\documentclass[english]{elsarticle}
\usepackage{amsmath}
\usepackage{amssymb}
\usepackage{amsfonts}
\usepackage{mathrsfs}
\usepackage{graphics}
\usepackage{lscape}
\usepackage{changebar}
\usepackage{graphicx}
\usepackage{epsfig}

\newcommand{\etal}{\textit{et al.} }

\newcommand{\del}{\mbox{\boldmath{$\nabla$}}}
\renewcommand{\vec}[1]{\mathbf{#1}}
\newcommand{\deriv}[2]{\frac{\partial #1}{\partial #2}}
\newcommand{\chib}{\mbox{\boldmath$\chi$}}

\newcommand{\fp}{\mathbf{F}^p}

\newcommand{\dpb}{\mathbf{D}^p}

\newcommand{\tb}{\mathbf{T}}
\newcommand{\mb}{\mathbf{M}}

\begin{document}
\begin{frontmatter}
\title{Nonlinear elasto-plastic model for dense granular flow}
\author{Ken Kamrin}
\ead{kkamrin@seas.harvard.edu}
\ead[url]{http://people.seas.harvard.edu/$\sim $kkamrin/}
\address{Harvard School of Engineering and Applied Sciences, 29 Oxford St, Cambridge, MA 02139}

\begin{abstract}
  This work proposes a model for granular deformation that predicts
  the stress and velocity profiles in well-developed dense granular
  flows. Recent models for granular elasticity (Jiang and Liu 2003)
  and rate-sensitive fluid-like flow (Jop $\etal$ 2006) are
  reformulated and combined into one universal elasto-plastic
  law, capable of predicting flowing regions and stagnant zones
  simultaneously in any arbitrary 3D flow geometry.  The unification
  is performed by justifying and implementing a Kr\"{o}ner-Lee
  decomposition, with care taken to ensure certain
  continuum physical principles are necessarily upheld. The model is
  then numerically implemented in multiple geometries and results are
  compared to experiments and discrete simulations.
\end{abstract}

\end{frontmatter}


\section{Introduction}

A general constitutive law for granular materials, valid over all
common regimes of deformation, remains an important open problem for
several communities. Mechanicians and materials engineers have modeled
granular materials for centuries guided by principles of continuum
solid mechanics ~\cite{nedderman,Sok,hill,
  wroth,anand00,savage98,kamrin07a, spencer64}, where various failure and yield
postulates have been employed.  In recent years, a resurgence of
interest in the topic has arisen among physicists
~\cite{jaeger96,degennes99,kadanoff99,edwards01,halsey02,aranson06},
where modeling efforts have been directed primarily toward statistical
and fluid dynamical approaches.

Some of the difficulty in modeling dense granular materials is due to
the existence of highly disparate responses when material is flowing
``slowly'' or ``moderately'' (see section \ref{regimes1}), both of
which only vaguely relate to the completely static response. To
describe certain phenomenological behaviors, an astounding array of
physical hypotheses have been suggested.  These include: rolling and
static phase interactions
~\cite{bouchaud94,bouchaud95a,boutreux98,boutreux99}, Bagnold rheology
generated by ``granular eddies''~\cite{ertas02}, granular
temperature-dependent viscosity~\cite{savage98}, density-dependent
viscosity~\cite{losert00,bocquet02}, non-local stress propagation
~\cite{mills99,bouchaud95}, free-volume diffusion opposing
gravity~\cite{lit58,mullins72,nedderman79,bazant06,rycroft06a},
``shear transformation zones'' coupled to free-volume
kinetics~\cite{lemaitre02,lemaitre02c}, and partial fluidization
governed by a Ginzberg-Landau order
parameter~\cite{aranson01,aranson02}. Such theories can typically fit
certain experimental data~\cite{midi04}, usually for a specific
geometry: avalanching surface flows
~\cite{bouchaud94,bouchaud95a,boutreux98,boutreux99,aranson01},
inclined planes ~\cite{bagnold54,ertas02}, Couette
cells~\cite{losert00,bocquet02}, inclined chutes
~\cite{pouliquen96,pouliquen01}, or wide
silo~\cite{lit58,mullins72,nedderman79,bazant06,rycroft06a}.  A good
review of granular flow theories can be found in \cite{alltheories}.

Some models have been developed that are general enough for multiple
flow environments \cite{anand00,wroth,kamrin07a,aranson02}.  In
particular, the work of Jop \etal \cite{jop06} introduces and tests a
generic, 3D, non-Newtonian fluid constitutive law representing
``moderate'' sized granular shear-rate as a local function of the
Cauchy stress.  The material is treated as a Bingham fluid and a
computation of the stress and flow profiles can be made in regions
above a Drucker-Prager yield criterion. While a breakthrough by many
standards, the law makes rigid-body assumptions in static phases where
stresses are not computed.  In many circumstances, granular assemblies
can establish static zones at walls (such as in flowing hoppers and
silos \cite{nedderman,Sok}), and effective flow computation is
impingent on the ability to model traction-based interaction laws
between sub-yield material and the system boundaries. Stress-based
quantities in sub-yield zones are needed to approach a variety of open
questions in granular flow theory, such as how ``kick-stresses''
\cite{nedderman,Dre} develop in draining storage bins.

This work seeks a general constitutive law for 3D, dense flowing
granular materials.  Specifically, we focus on flows with significant
regions of moderate shearing, as would be expected in day-to-day
environments like flowing chutes, hoppers, and heaps. To this end, the
model of Jop \etal is integrated with an appropriate granular solid
response, by way of a joint elasto-plastic treatment. The unified
formulation can be used to predict the stress and flow profiles
everywhere, for an arbitrary geometry with arbitrary admissible
kinematic/traction boundary conditions.  For sufficient generality and
accuracy in describing the static phase, we ultimately choose to pair
with the granular elasticity law proposed recently by Jiang and Liu
\cite{jiang03}.  This happens to be a natural choice for several
reasons, not the least of which is its proven capability to represent
several static and acoustic phenomena observed in granular matter.

The process of constructing a combined law is itself non-trivial. Both
components require some degree of reinterpretation--- the Jop \etal
law is converted to an evolution law for the plastic deformation
gradient, and the Jiang-Liu elasticity law is modified to permit
flowing states while maintaining elastic deformation.  Moreover, some
essential postulates of the chosen elasto-plastic framework are not
obvious in well-developed flow, requiring additional substantiation.

This paper is organized as follows. First, we describe some general
requirements for a continuum treatment of granular statics/flow
(section \ref{continuum}). Then a discussion of past solid and flow
constitutive laws for dense granular media are reviewed (sections
\ref{statics} and \ref{flow}), ultimately culminating in a description
of the two laws mentioned above. The process of combining these
responses is discussed (section \ref{physical}), which ends with a
summary of equations for the proposed model. Finally, section
\ref{numerical} provides discussion on the numerical implementation of
these equations and displays results in several flow geometries, with
comparisons against the data of \cite{jop06},\cite{midi04}, and
\cite{rycroft09}.

\section{Granular matter as a  continuum}\label{continuum}

Before development ensues, we take a moment to lay out the limits of
validity for a first-order continuum treatment of a discrete granular
collection, as shall be proposed herein. For a more detailed set
of arguments on deterministic limits in random composite materials,
the reader is directed to the seminal work of Ostoja-Starzewski
\cite{ostoja05}\cite{ostoja06}.

Recent Discrete Element Method (DEM) simulations of \cite{rycroft09}
have shown that in multiple 3D well-developed flow environments, a
dense granular element of width $\sim 5d$ (for $d=$ particle diameter)
captures many of the \emph{plastic flow} properties expected of an
RVE. Among neighboring volume elements of this width, the average
stress and deformation rate appear to vary smoothly.  Within a flowing
element, the eigenvectors of the instantaneous space-average Cauchy
stress align to a high degree with those of the deformation rate
tensor, evidencing the onset of a deterministic relationship between
the two fields. Moreover, a predictable dependence of the packing
fraction on the pressure and shearing rate emerges at this
length-scale. Of course, the representative element width is not a
universal constant; environments with large spatial gradients
necessitate smaller elements.  Thus the $5d$ element width can be
interpreted as a common limit for day-to-day flows, compacted
primarily by gravity, where boundary conditions on the global body
vary on a scale larger than $O(10d)$.


The result is not outlandish in light of the past observations by
Glasser and Goldhirsch \cite{glasser01} and Goldenberg $\etal$
\cite{goldenberg06}, where in depth 2D studies on disk arrays were
performed to quantify the effects of spatial averaging on granular
stresses. Rycroft's result is in line with the extremal trends
observed over the spectrum from rapid/dilute disk flows to
static/elastic deformation of a disk assembly.  Furthermore, it is
found that static materials gain resolution independence on a similar
scale to flowing materials, suggesting that a $5d$ element width could
give valid RVE behavior for dense granular materials in either static
or flowing states.

Previous arguments against a granular continuum, particularly in the
static phase but also during flow, have focused on the presence of
force chains in granular matter.  In 2D experiments and simulations of
disk assemblies \cite{behringer05,geng01}, concentrated chains of
interparticle force have been shown to exist over many particle
widths.  It was argued that if forces are not homogeneous at a
meso-scale, continuum relations at this scale cannot exist.  However,
in 3D simulations of flowing monodisperse sphere packings, we observe
that the force chains have a dramatically shorter length
\cite{rycroft09}.  A possible geometric explanation for this
phenomenon is that a 3D granular assembly has a much higher average
coordination number, reducing the likelihood that only two contacts
maintain the majority of the force on one grain.  These simulations
also include interparticle contact friction, which may contribute to
the smoothing out of force chains as has been previously shown in 2D
static disk assemblies \cite{goldenberg05}.  Typically speaking,
whether flowing or static, we observe that a $5d$ granular volume
element contains a diffuse network of contact forces, enabling a
sufficient degree homogenization of the stresses at this scale.

Frequently, a granular collection that has been set into motion will
possess regions of fluid-like flow adjacent to essentially solid-like
regions.  Examples of such flows include wide draining silos
\cite{choi05,samadani99} and hoppers (so-called ``core flow'') where a
broad column of material extending upward from the orifice flows like
a fluid, while regions closer to the side walls remain almost
completely static.  As will be discussed later, perfectly clear
solid/fluid interfaces are rarely observed in granular flow, which has
led some to believe that the solid-like zone is actually just a
`highly viscous' fluid region \cite{savage98,losert00,bocquet02}.
Solid-like material does undergo intermittent rearrangement events
when close to a zone of moderate flow-rate, but we find that the
stresses in these regions have essentially zero rate-dependence. For
example, when a DEM simulation of silo flow is suddenly arrested, say
by shutting the orifice, the stresses in the solid-like regions remain
virtually unchanged, supporting static shear stress like a solid.  It
appears that a mechanism disconnected from the flow-rate or any notion
of viscosity is responsible for maintaining the stress tensor in
solid-like granular matter, even if occasional failure events are
occurring within. This suggests that a continuum description for
granular statics is likely to be disjointed from the flow description.
Hence, we seek a general model that splices separate statics and flow
laws into one constitutive form.

\section{Continuum statics}\label{statics}
\subsection{Stress-only laws vs. elasticity}

To close the equations for static force balance, two approaches have
been utilized: stress-only laws and elasticity. Stress-only
relationships constrain the stresses directly, by asserting that the
stress tensor must satisfy some a priori relationship. Examples
include: Janssen's law of differential slices (originally proposed by
Janssen in 1895) where vertical and horizontal stresses are set to be
proportional, limit-state Mohr-Coulomb plasticity or the ``Incipient
Failure Everywhere'' (IFE) hypothesis \cite{Sok} where a Mohr-Coulomb
failure line is required to exist at all locations, and ``Oriented
Stress Linearity'' \cite{bouchaud95} where stresses propagate in
directions aligned with the presumptive microstructure of the packing.

While stress-only relationships are convenient and have had some
success, their physical assumptions can be questionable.  For example,
static granular matter is rarely in a limit-state of incipient failure
\cite{rycroft09} and wall shear is not compatible with a Janssen-style
stress tensor \cite{nedderman}.  Most stress-only laws are defined only for
2D media, which brings out issues of generality.  They often predict a
``hyperbolic'' character to the stress profile, where stress
quantities propagate in certain directions from the boundaries.  The
notion of force propagation was backed chiefly by the observation of a
double-peak pressure distribution beneath a bed of grains undergoing a
point force from the top. Work by Goldenberg and Goldhirsch
\cite{goldenberg05} has shown, however, that in the presence of
interparticle friction and a large system size to particle size ratio
(as commonly found in engineering applications) the pressure
distribution is indeed a broad single peak beneath the point force, as
one would expect for an elastic bulk media.

This brings us to elasticity, which shall be our preferred method for
granular statics.  It is a sensible approach seeing as the grains
themselves are elastic bodies presumably enabling generalization to an
Effective Medium Theory (EMT) where grain-level elasticity extends
statistically to a continuum mean-field theory
\cite{duffy57,digby81,walton87,norris97}.  Reversible (elastic)
deformations have been observed in granular matter for strains less
than $10^{-4}$ \cite{kuwano02}.  This is negligibly small compared to
the size of typical plastic deformation. However, grains are commonly
composed of stiff material indicating the important role that small
elastic strains may play in the development of the stress profile.  It
could be argued that the barely noticeable elastic strain of a static
assembly is what originally impelled scientists to seek stress-only
laws, so that rigid-body assumptions could be used.  Rather, it seems
there is no generally applicable stress-only constraint to accurately
define a 3D stress tensor. The elastic strains are small but
non-ignorable, and bear essential importance to the stress
description.

Bulk elasticity for cohesionless grains is not likely to have a simple
form since, for example, the material is unable to support tension and
thus the small strain response cannot be approximated as
linear. Nonlinear EMT has been derived from Hertz-Mindlin
interparticle contact mechanics \cite{duffy57,mindlin53} and modified
by others \cite{evesque98}, offering reasonable albeit not completely
satisfactory agreement with experiments \cite{kuwano02, goddard90}. A
recently proposed elasticity model encompassing many of the same
features as EMT was proposed by Jiang and Liu \cite{jiang03} in 2003.
This state-of-the-art formulation has had multiple successes and is
well-suited to our end goal of combining with a plasticity model.

\subsection{Effective Medium Theory for  bulk granular elasticity}\label{EMT}

In the classical work of Hertz, two perfectly elastic spheres that are
pressed into contact repel each other with a contact force that
depends on the radii of the spheres and the ``apparent overlap''.  For
two spheres of identical radius $R$ located at $\vec{x}_1$ and
$\vec{x}_2$, the normal force contact law is

\begin{equation}\label{hertz}
F_n(\delta)=\frac{2}{3}k_nR^{1/2}\delta^{3/2}
\end{equation}
where $\delta$ is the normal overlap
$(1/2)\left(2R-||\vec{x}_2-\vec{x}_1||\right)$.  The parameter $k_n$
is an effective stiffness equal to $4G_g/(1-\nu_g)$ where $G_g$ and
$\nu_g$ are the shear modulus and Poisson ration of the sphere
material.

Presume a granular collection with an isotropic distribution of
contacts that undergoes affine deformation when stressed. The Hertzian
contact law generalizes to an EMT that gives the following bulk
modulus under pure compression:
\[\kappa=\frac{1}{18\pi}\Phi Z k_n (-\frac{1}{3}\text{tr}\vec{E})^{1/2}\]
where $\Phi$ is the volume fraction, $Z$ is the mean coordination
number, and $\vec{E}$ is the (infinitesimal) strain tensor.  The
nonlinearity can be seen as arising directly from the nonlinearity of
Hertz's contact law. The major point is that the bulk modulus scales
with (compressive strain)$^{1/2}$, or equivalently as $p^{1/3}$ under
isotropic compression.  This has been verified directly in large-scale
DEM simulations of compressed sphere packings both with and without
interparticle contact friction \cite{makse04}.

Supposing frictionless spheres, a mean-field shear modulus can also be
derived in a similar fashion by analyzing an arbitrary affine
deformation including shear strain. One ultimately finds that the bulk
and shear moduli under non-zero shear scale similarly to the above form:

\begin{equation}\label{moduli}
\kappa\propto G\propto \Phi Z (-\text{tr}\vec{E})^{1/2}\propto(\Phi Z)^{2/3}p^{1/3}
\end{equation}

The inclusion of shear strain renders EMT less accurate. The moduli
agree with the above scalings but only under low pressures ($<$10 MPa)
\cite{makse04}.  Experiments verify that the shear dependence on $p$ is
characterized by a larger exponent at higher pressures
\cite{goddard90}.  This could be because shearing under higher
pressures tends to render the affine displacement assumption less
valid. Time-dependent relaxation occurs, which significantly
complicates a determination of the shear modulus. Attempts to improve
the EMT elastic moduli during shear are not vastly affected by
including tangential forces in the analysis; the inclusion of
Mindlin's tangential force law \cite{mindlin53} between spheres does
not change the above scalings.

\subsection{The Jiang--Liu granular elasticity law} 
In 2003, Jiang and Liu proposed a modified approach to granular
elasticity \cite{jiang03}.  Rather than continue laboring on a
microscopically derivable mean field theory, their aim was to
augment the more successful results of EMT with presumptive forms that
capture known macroscopic behavior.

They consider an elastic free energy density of the following form:
\begin{equation}\label{psi}
\psi(\vec{E})=B\sqrt{\Delta}\left(\frac{2}{5}\Delta^2+\gamma^2/\xi\right)
\end{equation}
where $B$ is a relative stiffness that can vary with packing fraction. The
compressive strain and shear strain are measured respectively by
\[\Delta=-\text{tr}\vec{E} \ , \ \ \gamma=|\vec{E}_0|\]
where for any tensor $\vec{A}$, we adopt the notation
$\vec{A}_0\equiv\vec{A}-(1/3)\text{tr}\vec{A}$ and
$|\vec{A}|\equiv\sqrt{\sum_{i,j}A_{ij}^2}$.  The dimensionless
parameter $\xi$ will be discussed shortly.  In the small displacement
limit this gives the Cauchy stress
\begin{equation}\label{elastic_stress}
\vec{T}=\deriv{\psi}{\vec{E}}=2\underbrace{\frac{B\sqrt{\Delta}}{\xi}}_{G(\Delta)}\vec{E}_0+\underbrace{B\sqrt{\Delta}\left(1+\frac{\gamma^2}{2\xi\Delta^2}\right)}_{\kappa(\Delta, \gamma)}(\text{tr}\vec{E})\vec{1}.
\end{equation}
Under isotropic compression, the pressure is proportional to
$\Delta^{3/2}$, in agreement with the most successful result of EMT.
The shear modulus scales with $\Delta^{1/2}$ as in EMT, but the added
nonlinearity in the full form of the bulk modulus allows for some
important properties not well-represented by EMT.

In an arbitrary state of strain $(\gamma,\Delta)$, the corresponding Drucker-Prager friction state is
\[\mu=\frac{\tau_{eq}}{p}=\sqrt{2}\left((\gamma/\Delta)^{-1}\xi + (\gamma/\Delta)0.5\right)^{-1}\]
By varying the input $r\equiv \gamma/\Delta$, one can show using basic
calculus that $\mu$ has a global maximum value of $1/\sqrt{\xi}$.
Thus, by selecting $\xi$ accordingly, we can prevent certain states of
friction from ever arising.  This is one of the defining features of
Jiang-Liu elasticity.

\begin{figure}
\begin{center}
\includegraphics[width=2.5in]{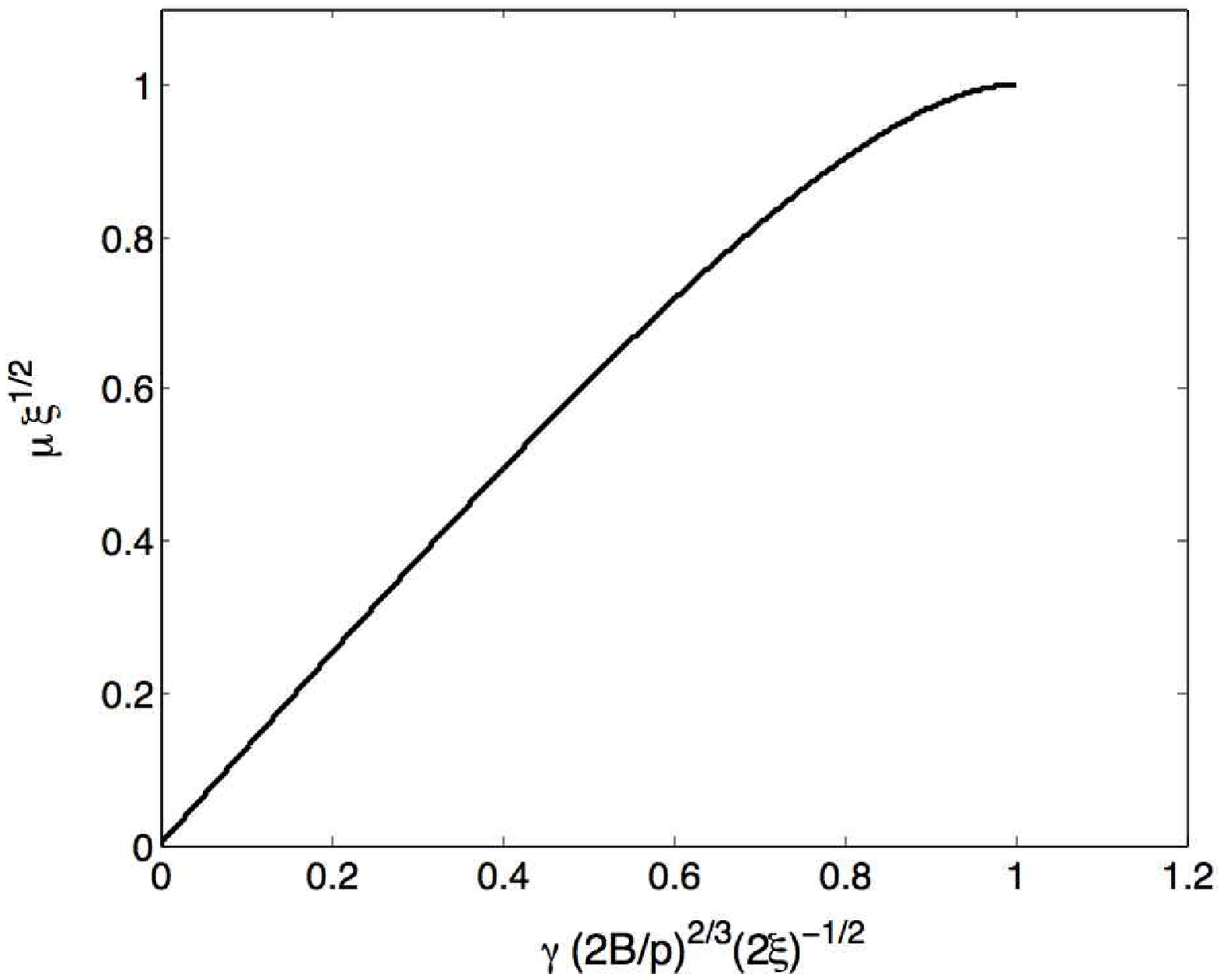}
\ \ \ \ \ \ \ \ \ \ \includegraphics[width=2.3in]{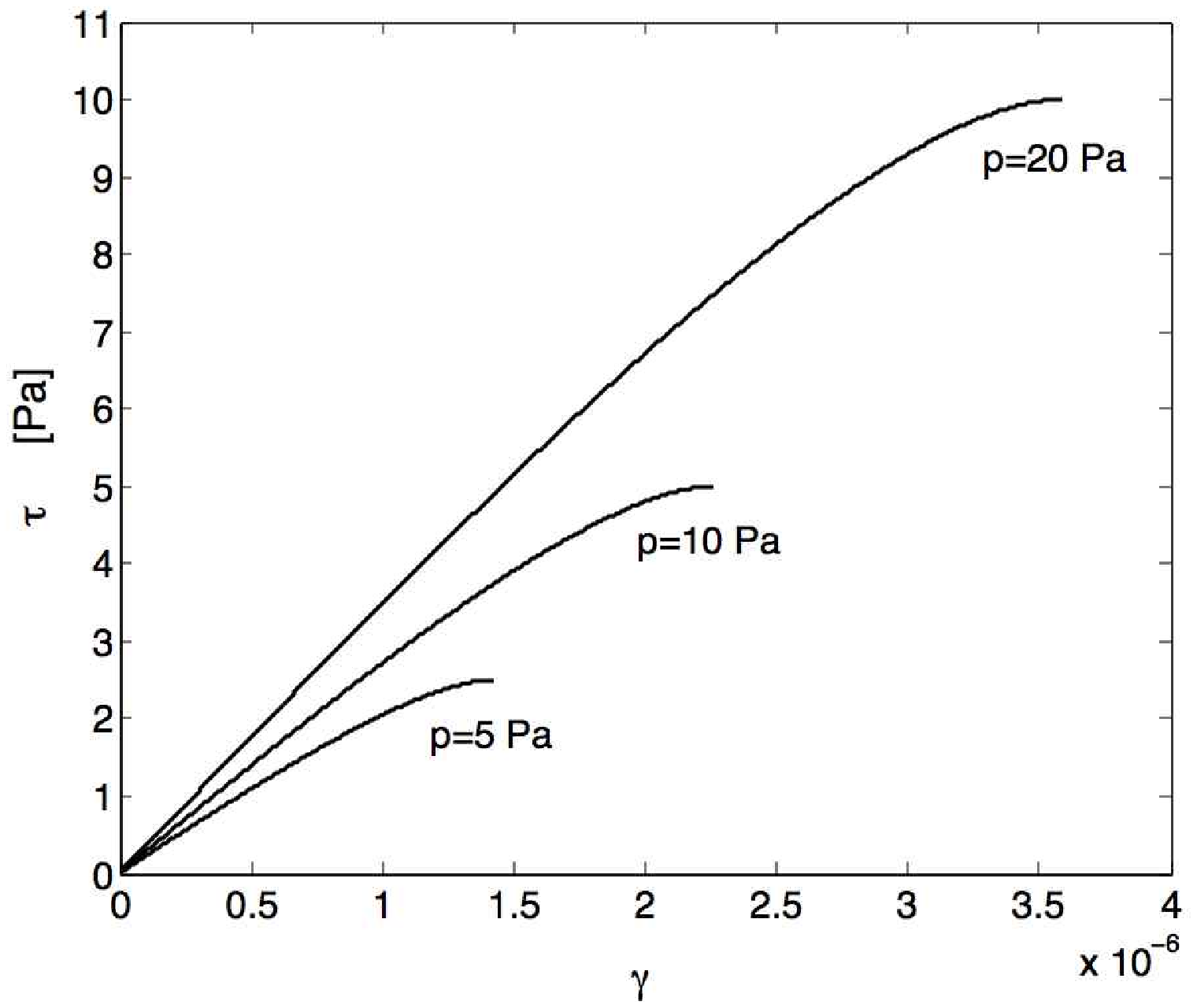}
\caption{The shear stress vs. shear strain relation under Jiang--Liu
  elasticity represented as a single dimensionless plot (left), and
  plotted equivalently as a family of shear stress vs. shear strain
  curves in SI units using $\xi=4$, $B=7\times 10^9$ Pa, each curve determined by the
  applied compressive pressure (right).} \label{elast_plot}
\end{center}
\end{figure}

In the work of Jiang and Liu, static materials were the primary
concern.  Hence, a static yield criterion $\mu_s$ was declared and
$\xi$ was set to $\mu_s^{-2}$ so as to require that no purely elastic
state can exist above $\mu_s$.  The work herein shall attempt to
integrate elasticity within a complete elasto-plastic framework, so
the basis for selecting $\xi$ will be slightly different. At the point
where the stress state reaches $\mu_{max}$, the elasticity relation
loses convexity. The particular condition for convexity is
$|\vec{E}_0|\le\left(p/2B\right)^{2/3}\sqrt{2\xi}$. Figure
\ref{elast_plot} elucidates the fashion by which $\mu$ increases to
its maximum under increasing shear strain.  Plots are discontinued in
non-convex regions, where the elasticity law is deemed void.

More than just a model that connects with yield, the Jiang--Liu
elasticity model has had some convincing successes in representing
granular statics:
\begin{enumerate}
\item The nonlinear form of $\vec{T}$ produces a stiffness tensor
  $\deriv{T_{ij}}{U_{kl}}$ that agrees to a satisfactory extent with
  the form of the stiffness tensor extracted from experimental data
  \cite{kuwano02}.
\item The model predicts Janssen-type saturation of the wall stresses
  in a tall silo.  The ratio of vertical to horizontal compressive
  stresses in the silo is found to be approximately constant when not
  close to the walls.  This verifies the commonly used notion of a
  ``coefficient of redirection'', which has been verified in DEM
  simulations \cite{rycroft09}.
\item The model predicts that a granular material under simple shear
  stress responds anisotropically to the addition of a point-load at a
  surface. Such anisotropy under pre-stress is a well-known granular
  phenomenon that is captured appropriately in the nonlinearity of the
  Jiang--Liu model.
\item It has been observed that preparation history is largely
  responsible for the stress dip that one often observes under the
  peak of a sand pile. This fact is reproduced by the elasticity model
  when solved assuming an initial packing fraction distribution that
  one might expect for a conical pile constructed from grains flowing out of a
  nozzle.
\end{enumerate}

\section{Continuum flow}\label{flow}
\subsection{Bagnold scaling and relevant dimensionless quantities}\label{bagnold}
We now move on to a discussion of granular flow laws for inelastic
deformation. Bagnold's seminal work on granular flow followed from
shear experiments on granular/fluid suspensions.  In the
``grain-inertia'' regime where the effects of the interstitial fluid
are small, Bagnold found that both shear and normal stresses on the
shearing wall depend quadratically on the shear rate \cite{bagnold54},
a phenomenon that came to be known as ``Bagnold scaling''.  Bagnold
scaling has been verified for dry grains in a number of experiments
and simulations
\cite{lois05,silbert01,pouliquen99,prochnow_thesis,dacruz05}.  An
explicit form in the case of simple shear of a dense configuration of
dry grains can be expressed as the following pair of dimensionless
relations:

\begin{eqnarray}
&&\Phi=f(I) \ \ \text{for} \ I=\frac{\dot{\gamma}d}{\sqrt{P/\rho_s}}\label{I_Phi}
\\
&&\mu=g(I) \ \ \text{for} \ \mu=\frac{\tau}{P} \label{mu_Phi}
\end{eqnarray}
In the above, $\Phi$ is the packing fraction, $P$ is the pressure on
the shearing plate, $\tau$ is the shear stress, and the steady shear
rate is $\dot{\gamma}$. The dimensionless number $I$ is commonly
referred to as the inertial number or normalized flow rate, and $\mu$
is the effective friction.

The simplest way to understand these equations is through dimensional
analysis. The major physical quantities involved in a gravity-free
simple shearing between long rough plates are the material parameters
$d$ and $\rho_s$, and the variable quantities $\Phi$, $\tau$, $P$, and
$\dot{\gamma}$.  This tacitly ignores the possibility of any other
length-scales playing a role and presumes that collisions are fully
dissipative (pressure high enough to damp out restitution), two
assumptions whose consequences are important and will be discussed
shortly.  The particle-on-particle contact friction $\mu_p$ is also
ignored. Granted, $\mu_p$ does affect $\mu$, but it has been found to
merely translate the $\mu$ vs. $I$ relationship vertically
\cite{dacruz05}.

With these assumptions, the packing fraction and effective friction in
a simple shear apparatus should arise uniquely from $P$ and
$\dot{\gamma}$ as a matter of cause and effect. Nondimensionalization
implies that $\mu$ and $\Phi$ should depend only on $I$.  Unlike a
Newtonian fluid where a temperature time-scale exists, the quadratic
dependence of stress on shear rate can be seen to arise from the fact
that $\dot{\gamma}$ can only be scaled by the square root of a
stress quantity.

\subsection{Flow regimes}\label{regimes1}

It is important to clarify the properties and overall validity of
equations \ref{I_Phi} and \ref{mu_Phi} over the range of possible flow
behaviors.  First, zero collisional restitution was presumed in
justifying these equations.  The assumption is valid if the kinetic
energy of a collision is always dissipated in full on impact,
presumably in the form of heat and sound.

For faster shear rates, this source of energy loss causes notable
rate-sensitivity. Resultantly, $I$ becomes one-to-one with $\mu$, and
$\dot{\gamma}$ is immediately determined by $\tau$ and $P$ as in a
non-Newtonian fluid. Gathering these properties into a general
classification scheme, a flow rate is deemed to be \emph{moderate}
when $I$ is large enough for rate-dependence, but small enough for the
flow to stay dense as per the collisional collapse argument. Data of
da Cruz $\etal$ \cite{dacruz05}, would suggest this regime lies within
the band $10^{-3}<I<10^{-1}$.  In day-to-day terms, the flowing region
of an hourglass is typically in the moderate range.

Moderate flows have the property of \emph{shearing dilation}, where
increasing the normalized flow rate causes the steady-state packing
fraction to decrease (i.e. $f$ in equation \ref{I_Phi} becomes a
decreasing function).  This should not be confused with \emph{shear
  dilation}, which refers to the drop in packing fraction as a
function of total shear that occurs to a dense assembly at the
beginning stages of a shear deformation.  Flows too slow to be deemed
moderate may still undergo shear dilation due to packing geometry, but
rate effects like shearing dilation only set in for faster flows.

\begin{figure}
\begin{center}
\epsfig{file=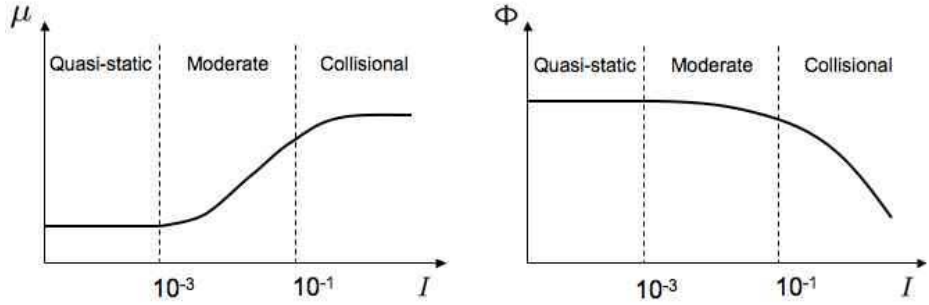, width=5in, clip} 
\caption{Qualitative diagrams (primarily due to \cite{dacruz05})
  showing the variation of dimensionless parameters through the
  various flow regimes under simple shearing. In other geometries, the
  quasi-static regime of $\mu$ is not as clear to define in terms of
  $I$ due to meso-scale effects. The coefficient of restitution
  affects $\Phi$ vs. $I$ in the collisional regime. The moderate
  regime is relatively well-determined as a sole function of
  $I$.}\label{schematic}
\end{center}
\end{figure}

Beyond the moderate range, dilute or \emph{collisional} flows occur in
general for $I>10^{-1}$ and correspond to the breakdown of the zero
restitution assumption. When $I$ becomes this large, particle
collisions are accompanied by some additional ``bounce-back'' akin to
a gas. The collisions are chiefly binary, and particles rarely
maintain long lasting contacts. These flows may require a temperature
quantity to store fluctuational energy and can be described with
dissipative Boltzmann kinetics.  A sand blast could be considered a
basic example of the collisional regime.

On the other side of the spectrum, where $I<10^{-3}$, we enter the
\emph{quasi-static} regime.  The packing fraction in simple shear does
not vary noticeably with $I$ in the (time-averaged) steady limit---
the inertial time is always small enough for the particles to find
tight compaction.  Without a significant contribution from collisional
dissipation, rate-dependence subsides, and more complicated
dissipation mechanisms dominate such as frictional contact sliding.  The
stress/strain-rate relationship becomes singular; driving the system
with a range of quasi-static normalized shear rates all give the same
time-average value for $\mu$.  Soil creep is typically in the
quasi-static range.

The discussion thus far has focused on simple shear, where stresses
and time-average flow are spatially uniform. While the other regimes
display a strong local rheology, quasi-static flows are sensitive to
gradients in the fields as expressed through some non-local term with
an associated length-scale \cite{midi04,bazant06,kamrin07a,aranson02}.  For
example, consider steady flow in an annular Couette cell. Slow moving
material far from the inner wall is observed to constantly creep
\cite{midi04}, even though the stress state should be below yield by
all common local measures.  Indeed, the motion appears to be caused by
a non-local effect where faster flow near the inner wall has
effectively ``bled out'' into neighboring material. Grain-level
specifics such as roughness, grain shape, and configurational
statistics (including wall effects) should affect the non-local flow
behavior through the new length-scale.  The size, dynamics, and
general interpretation of the length-scale are object of debate,
though most agree its size should be on the order of several particle
widths.

Observe figure \ref{schematic} for a schematic view of how $\mu$ and
$\Phi$ vary throughout the flow regimes (in simple shear). The major
points of this discussion are: moderate flow is much simpler than
quasi-static, and collisional flow is outside our interest as it is
not dense.  Moderate flows are characterized by a local rheology
relating $I$ to both $\mu$ and $\Phi$.  The regime is definitively
rate-dependent and equation \ref{mu_Phi} inverts into a fluid-like law
wherein the flow rate can be determined uniquely from the stress
state. As $I$ decreases to the quasi-static limit, however, quantities
that were previously ignorable become important.  Grain motion is
locally correlated at some length-scale causing nonlocality and a
larger role of grain-level properties. The flow does not permit a
fluid-like treatment as before, since the dissipation is largely
rate-independent.

\subsection{Quasi-static flow models}\label{regimes2}

To review dense flow models, we begin with those aimed at quasi-static
behavior. Since the stresses are not a direct function of the flow
rate, they are typically obtained by asserting an elasticity law upon
a small component of the total deformation.  Linear elasticity is
often presumed, but this may be an oversimplification that carries
more consequences at lower stresses as described in section
\ref{statics}.  Critical State Theory \cite{wroth},
Rudnicki--Rice-type modeling \cite{rudnicki75}, the Anand--Gu model
\cite{anand00}, and the model of La Ragione \etal \cite{ragione08}
each can be used for granular deformation in this fashion. Other
models couple to the IFE stress formulation in 2D, such as
rate-independent coaxiality \cite{nedderman} and the Stochastic Flow
Rule \cite{kamrin07a}.

Some quasi-static theories account for nonlocality, though none have
been proven to do so reliably for steady flow in arbitrary 3D
environments. Several theories are based on new definitions of temperature,
which introduce a length-scale via a ``heat equation''. These include
Shear Transformation Zone (STZ) Theory \cite{lemaitre02, falk98},
which relies on an effective temperature governing STZ creation, the
dense flow theory of S. B. Savage \cite{savage98}, which defines a
granular temperature to measure strain-rate fluctuations, and Edwards
statistics \cite{edwards91}, which utilizes a temperature-like
`compactivity' derived from an entropy per free volume. These models
provide interesting physical insight, but do not appear to be at the
point of development that simulating arbitrary flows would be
possible--- some are restricted to 2D, the boundary conditions for the
new temperature are rarely obvious, and the equations may not be
closed except in a few symmetric test cases. Besides temperature
approaches, nonlocal behavior could also be described with a diffusing
order parameter, as in \cite{kamrin07a,aranson02}, or through a more
general strain-gradient plasticity theory
\cite{gurtin05a,gurtin05b,vardoulakis91,hattamleh04,hashiguchi07}.

The quasi-static flow regime, though important, appears at the moment
too difficult to account for appropriately within a simple 3D
continuum framework. Likewise, the following concession is enacted:
\textbf{The model to be constructed in this work shall neglect
  quasi-static flow behavior altogether, opting instead to combine a
  static response directly with moderate flow
  rheology.}  
The model herein may ultimately serve as the backbone for a fuller
model that also incorporates the dependence of a length-scale on the
slow dynamics.  This possibility shall be considered in more depth
when we compare predictions of the model directly to experimental/DEM
data. But for now, we accept inaccuracy in describing quasi-static
motion in exchange for a closed, general model capable of giving
worthwhile predictions over the full range of dense material behavior,
accounting for both statics and flow.

\subsection{Moderate flow law of Jop \etal}\label{jop}

A closed form law to predict moderate flow must now be discussed. Since
the moderate flow regime is (monotonically) rate-dependent, the
function $g$ in equation \ref{mu_Phi} should be invertible.  Hence,
increasing the normalized shear rate $I$ results in higher $\mu$.
This notion may seem counterintuitive to the fundamental idea that
$\mu$ should \emph{decrease} to a kinetic value as the rate of sliding
picks up.  But recall, for moderate flow rates, the impact dissipation
dominates sliding affects.  In slower or transient flows, shear
weakening is indeed observed for ``overconsolidated'' material, and
accounted for in various models via hardening parameters
\cite{anand00,wroth}.

Based on results from numerical simulations of planar shear
\cite{dacruz05,iordanoff04}, and successful extensions to
plane-strain inclined chute flows \cite{midi04,silbert01}, the
experiments of \cite{jop05} were conducted to quantify $g^{-1}$ for
glass beads:

\begin{equation}\label{rheology}
I=g^{-1}(\mu)=I_0\frac{\mu-\mu_s}{\mu_2-\mu} \ \ \text{for} \ \mu>\mu_s.
\end{equation}
The values of the parameters were measured at $I_0=0.279$, $\mu_s=\tan
20.9^{\circ}$, and $\mu_2=\tan 32.76^{\circ}$. The relation states
that the normalized shear rate $I$ increases as the material is
sheared with higher $\mu$. But $\mu$ must exceed some static yield
value $\mu_s$ before any plastic flow ensues. There is also some
maximal $\mu$ value, $\mu_2$, and all steady shear flowes should be
tenable for a value of $\mu$ less than $\mu_2$.  Consequently, if
applied stresses exceed $\mu_2$ the flow is deemed accelerative with
no apparent steady state.  While this last point may be somewhat of an
oversimplification from a physical point of view, the approximation is
helpful especially in light of the fact that $\mu$ for dense flow is
usually well below $\mu_2$.

\begin{figure}
\begin{center}
\epsfig{file=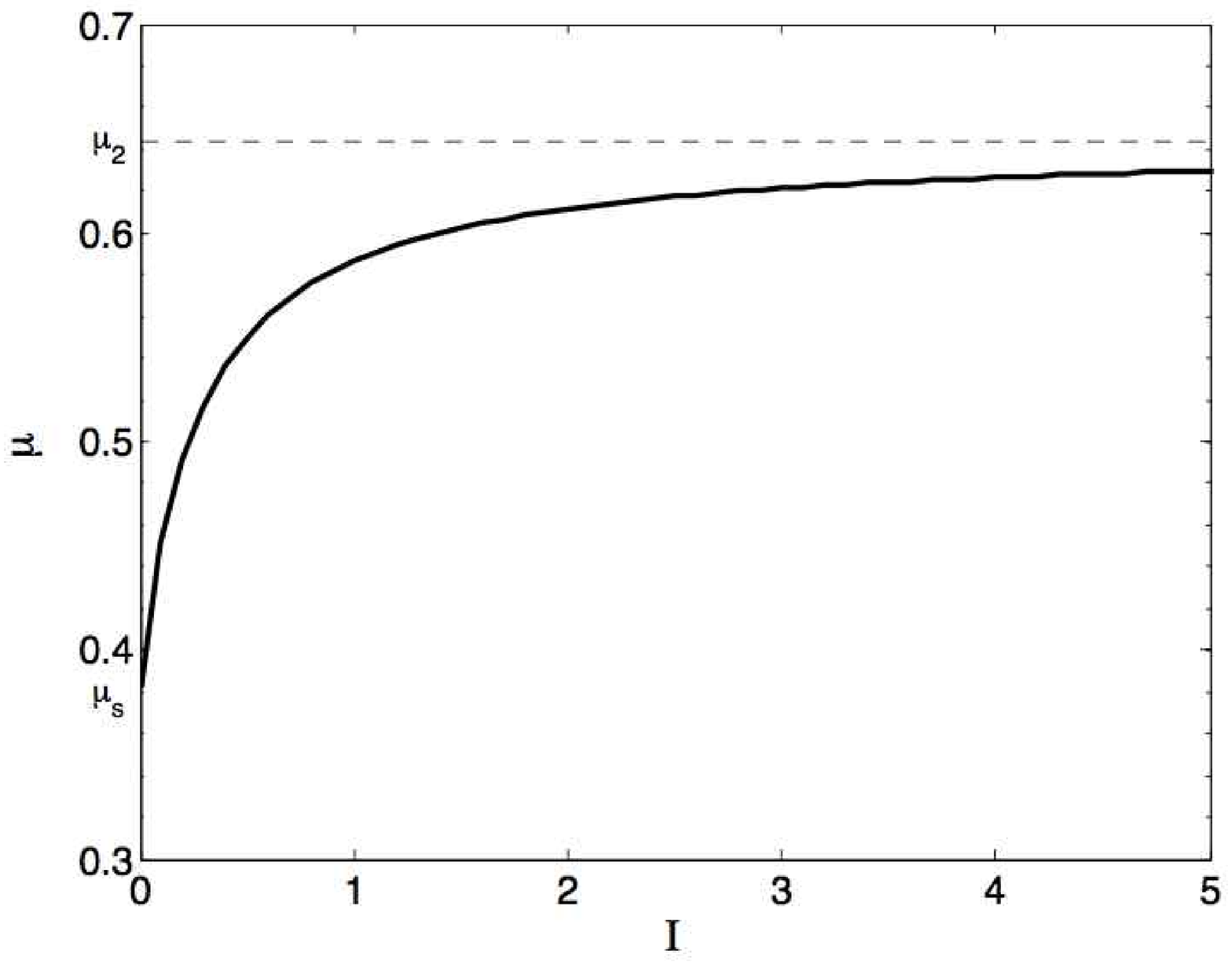, width=2.4in, clip} \ \ \
\epsfig{file=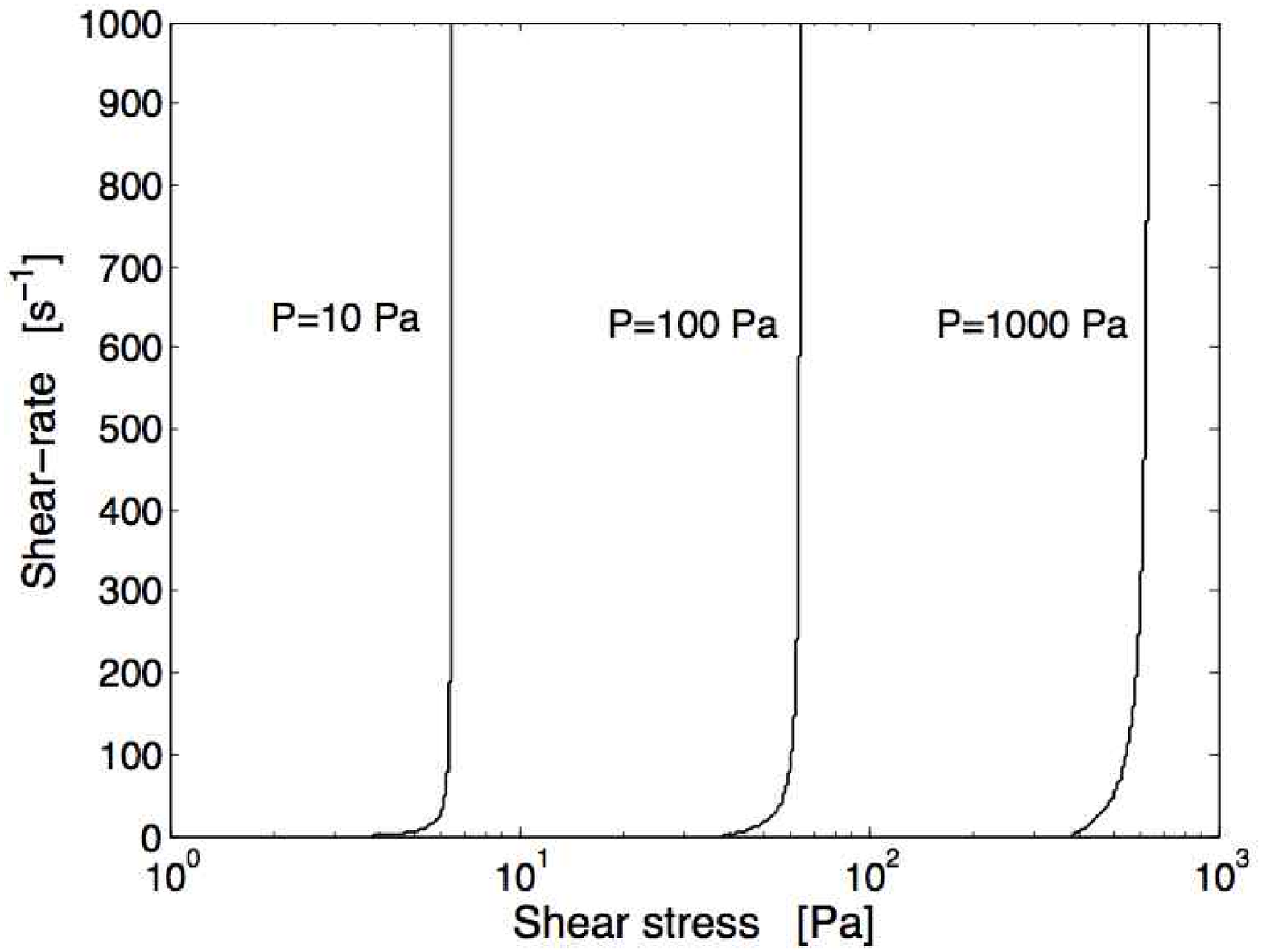, width=2.4in, clip}
\caption{The plastic flow rheology in simple shear plotted as one
  dimensionless relationship (left), and plotted equivalently as a
  family of shear-rate vs. shear stress curves in SI units, each curve
  determined by the applied compressive pressure (right).}\label{flow_rule_fig}
\end{center}
\end{figure}

The first attempt to extend equation \ref{rheology} into 3D was met
with high success. In Jop $\etal$ \cite{jop06},
\emph{codirectionality} was applied, presuming that the deformation
rate tensor
\[\mathbf{D}=(1/2)\left(\del\vec{v}+(\del\vec{v})^T\right)\]
is proportional to the deviatoric stress tensor
\[\vec{T}_0=\vec{T}-(1/3)(\text{tr}\vec{T})\vec{1}.\]
Written in full, Jop proposed the following  generalization of
equation \ref{rheology}:
\begin{equation}\label{flow_rule}
\vec{D}\frac{d}{\sqrt{P/\rho_s}}=I_0\frac{\mu-\mu_s}{\mu_2-\mu}\frac{\vec{T}_0}{\tau}
\end{equation}
where now $P=-(1/3)\text{tr}\mathbf{\vec{T}}$ and $\mu=\tau/P$ where
$\tau=|\vec{T}_0|/\sqrt{2}$ is the equivalent shear stress. When
$\mu<\mu_s$, we take $\vec{D}=\vec{0}$ establishing a Drucker-Prager
yield criterion.  Since the flow condition being used is
codirectionality, the flow rule is thus non-associative.

Since $\vec{D}$ is proportional to a deviatoric tensor, the flow rule
asserts plastic incompressibility.  While, as previously noted,
dilation in dense flow does occur, it is typically on the order of
only a few percent and quickly reaches a steady value over large
deformations.  Moreover, $\Phi$ is unnecessary to compute the
\emph{steady} shear rate, hence the approximation of plastic
incompressibility should have negligible effect on the velocity field
of a dense steady flow.  Note that evolution laws for the packing
fraction have not yet been quantified in this context. Some
quasi-static flow models attempt this \cite{anand00, wroth,
  rudnicki75} but rather than try to modify one of these, we go along
with the presumptions of Jop \etal and ignore plastic
dilation.  The bigger impact of this assertion is not on the flow, but
rather on the stresses in the static regions--- If plastic flow
transiently passes through a region that is ultimately static in the
steady-state, the dilation that occurred there affects the local
elastic moduli.

The flow model was tested experimentally in a chute
apparatus with rough sidewalls and bottom to induce non-trivial 3D
flow. Results were shown to match experiments to a high degree (always
within $15\%$) even while varying several parameters (e.g. inclination
angle, flow height and width). With positive experimental validation,
the flow model became one of the first to describe well-developed,
inhomogeneous, 3D, dense granular flows under moderate flow rates.

We now proceed to unite the Jop \etal plasticity law (equation
\ref{flow_rule}) with the Jiang--Liu elasticity law (equation
\ref{elastic_stress}).

\section{Combining elasticity and plasticity}\label{physical} 

\subsection{Kinematics}
First, some kinematic quantities of interest to the discussion shall
be briefly reviewed (see \cite{gurtin81} or \cite{anand04} for more
details). For a material element that begins at location $\vec{X}$ and
resides at $\vec{x}$ some time $t$ later, the \emph{motion function}
$\chib$ is defined by $\vec{x}=\chib(\vec{X},t)$. The elastic and
plastic responses are combined using the Kr\"{o}ner-Lee decomposition
of the deformation gradient
$\vec{F}\equiv \partial\vec{\chib}(\vec{X},t)/\partial\vec{X}$
\cite{kroner60,lee69}:
\begin{equation}\label{kroner}
\vec{F}=\vec{F}^e\vec{F}^p
\end{equation}

The decomposition models the deformation of an element at any time as
a progression of two stages of deformation.  First, the unstressed
\emph{reference} configuration is deformed by the plastic deformation
$\vec{F}^p$ to an \emph{intermediate} configuration, which is still
deemed as stress-free. From there, the material deforms
elastically via $\vec{F}^e$ to its final or \emph{deformed}
configuration. The intermediate configuration can be thought of as the
residual state that occurs when all stresses on a deformed volume element are
released, thereby unloading all elastic mechanisms, leaving only the
total plastic deformation.

The spatial velocity gradient can be expressed in terms of $\vec{F}$ via
\[\vec{L}\equiv\deriv{\vec{v}}{\vec{x}}=\dot{\vec{F}}\vec{F}^{-1}\]
where $\dot{}$ always represents the material time
derivative. Generalizing this, the elastic and plastic velocity
gradients are defined as
\begin{align}
\vec{L}^e&=\dot{\vec{F}}^e\vec{F}^{e-1}
\\
\vec{L}^p&=\dot{\vec{F}}^p\vec{F}^{p-1} \label{Lp}
\end{align}
which in turn enables definitions of the elastic and plastic deformation rate and the elastic and plastic spin:
\begin{align*}
\vec{D}^e&=\text{sym}(\vec{L}^e) \ \, \ \ \ \vec{D}^p=\text{sym}(\vec{L}^p)
\\
\vec{W}^e&=\text{skw}(\vec{L}^e) \ \, \ \ \ \vec{W}^p=\text{skw}(\vec{L}^p).
\end{align*}

The deformation gradient and its elastic/plastic counterparts are
presumed to have positive determinant, permitting the polar
decomposition
\begin{equation}\label{polar}
\vec{F}^e=\vec{R}^e\vec{U}^e
\end{equation}
where $\vec{U}^e$ is the elastic right stretch (which is symmetric
positive definite) and $\vec{R}^e$ is the elastic rotation. The
plasticity model we wish to use is incompressible and as such we have
$\text{det}\vec{F}^p=1$.  Consequently,
$\text{det}\vec{F}^e=\text{det}\vec{F}$.  From the elastic stretch, we
invoke the material Hencky strain measure to define the elastic
strain:
\begin{equation}\label{hencky}
\vec{E}^e=\log(\vec{U}^e)
\end{equation}

\subsection{Elasto-plastic constitutive picture}
The constitutive framework to be used herein is analagous to a
spring/damper combination in series, with Jiang--Liu elasticity
representing the spring deformation and the flow law of Jop \etal
representing the damper (with a yield criterion). Notably, the
stresses are modeled as being determinable everywhere from the field
$\vec{E}^e$.  When the resulting stresses satisfy yield, the plastic
flow law determines the flow-rate $\vec{L}^p$.

It is a non-trivial physical assertion to model all stresses in a
dense flowing granular material as being derivable from the elastic
deformation of the grains. Other microscopic stress agencies exist
such as internal viscous damping and fluctuational stresses. To
validate our assertion, discrete particle simulations of Rycroft \cite{rycroft_misc} were performed in several dense,
inhomogeneous flow environments. The instantaneous stress tensor over
an element is represented by the particulate formula
\begin{equation}\label{discrete_stress}
  \vec{T}=\frac{1}{V}\left(\sum_{i<j}^N\vec{r}^{(ij)}\otimes\vec{f}^{(ij)} \ - \ \sum_{i=1}^N m^{(i)}(\vec{v}^{(i)}-\bar{\vec{v}})\otimes(\vec{v}^{(i)}-\bar{\vec{v}})\right)
\end{equation}
for $N$ the number of grains in the element, $V$ the element volume,
$\vec{f}^{(ij)}$ the contact force of particle $i$ on particle $j$,
$\vec{r}^{(ij)}$ the vector connecting the centroids of particles $i$
and $j$, $m^{(i)}$ the mass of particle $i$, $\vec{v}^{(i)}$ the
velocity of particle $i$, and $\bar{\vec{v}}$ the average velocity
over all $N$ particles. The first term inside the parenthesis
represents a solid-like stress response derived from contact forces,
while the latter term is gas-like, accounting for stresses due to
velocity fluctuations. Rycroft found that the gas-like term is always
exceedingly small compared to the solid term.  The contact law used in
the simulation was visco-elastic (plus sliding friction) giving the
force decomposition
$\vec{f}^{(ij)}=\vec{f^e}^{(ij)}+\vec{f^v}^{(ij)}$. By comparing
the viscous and elastic force contributions, Rycroft found that
the elastic contribution is always vastly dominant.  Altogether,
\[\vec{T}\cong\frac{1}{V}\left(\sum_{i<j}^N\vec{r}^{(ij)}\otimes\vec{f^e}^{(ij)}\right).\]
This evidence suggests the elastic mechanism accounts for virtually all
stress in a flowing or static element of dense granular material,
lending reasonable validation to our proposed elasto-plastic
treatment.

\subsection{Modifying the Jiang--Liu elasticity law}

The Jiang--Liu elasticity law requires some slight modifications in
order to be used for elasto-plasticity. Since flowing materials can
have $\mu$ above $\mu_s$, the elastic law must permit these stress
states. Hence, we propose the following important modification to
Jiang--Liu elasticity: Set $\xi=\mu_2^{-2}$ instead of
$\xi=\mu_s^{-2}$.  Little loss in representing elasticity occurs from
this change. The $\xi$ parameter was engendered to represent the
macroscopic repose angle of a static granular assembly; it was not
determined from any quantitative microscopic requirements.

With $\xi=\mu_2^{-2}$, there can never be an elastic stress state that
has $\mu>\mu_2$. Looking back to equation \ref{flow_rule}, notice that
the flow rule cannot apply to a stress state that exceeds $\mu_2$.
Since the Jiang--Liu law admits a cap on the value of $\mu$, this
property prevents elastic stress states from entering the forbidden
regime of the Jop \etal flow law, which offers several benefits when
attempting to numerically solve for flow.  Analytically speaking,
capping the elastic stresses at $\mu_2$ is not necessary for a
solution; a cap above $\mu_2$ would also be acceptable, though the
plastic response would always preclude such states from arising.
Regardless, $\mu_2$ is a natural choice that minimizes the extent of
alteration of Jiang and Liu's original formulation.

\subsection{Mathematical specifics}\label{mathematical}
With the motivation provided, we now go about providing a
mathematically rigorous framework to unite the proposed elastic and
plastic responses. The following is based on a form for
thermodynamically compatible elasto-plasticity developed in
\cite{anand04}.  An abbreviated discussion shall be provided
here. More details can be found in \cite{kamrin_thesis}.

At the outset, Newton's equations of motion must be upheld for force
and torque balance, and mass is conserved:
\begin{equation}
\deriv{}{\vec{x}} \cdot \tb +\rho\mathbf{g} = \rho\dot{\mathbf{v}}, \ \ \ \ \ \  \tb = \tb^T, \ \ \ \ \ \ \rho =\rho_0(\text{det}\vec{F})^{-1}
\label{balance}
\end{equation}
where $\rho_0$ is the initial material density, equal to
random-close-packing (63\%) times $\rho_s$. To institute simultaneous
elastic and plastic constitutive responses, three major restrictions
are enforced
\begin{enumerate}
\item Frame-indifference
\item Non-violation of the second law of
thermodynamics
\item Isotropy of the reference and intermediate configurations.
\end{enumerate}


The presumption of isotropy, particularly in the intermediate state,
deserves some attention. Anisotropy of plastically deformed granular
material has been studied extensively both in 2D \cite{thornton06} and
in 3D where it has been quantified \cite{tsutsumi08} and modeled
\cite{zhu06,tsutsumi05} often by inclusion of a fabric tensor. These
studies show varying levels of anisotropy that evolve over a range of
strains that are large, but always significantly less then $50\%$. In
the limit of well-developed flows, however, experimental evidence of
\cite{tsai03} and discrete simulations of \cite{depken07,rycroft09}
support the use of an isotropic flow law.

With the presumed $\vec{F}^e\vec{F}^p$ decomposition, one can show via
power conjugacy that the stress interacts with the elastic and plastic
mechanisms through a surrogate known as the \emph{Mandel stress}
\begin{equation}\label{mandel}
\vec{M}=J^{e}\vec{F}^{eT} \tb \vec{F}^{e}.
\end{equation}
To connect to more common stress measures, $\vec{M}$ is equivalent to
$\vec{C}^e\vec{T}_{\text{II}}$ where
$\vec{T}_{\text{II}}=J^e\vec{F}^{e-1}\vec{T}\vec{F}^{e-T}$ is the
second Piola stress as measured using $\vec{F}^e$ instead of
$\vec{F}$, and $\vec{C}^e=\vec{F}^{eT}\vec{F}^e$ is the elastic right
Cauchy-Green tensor. As in the analogy with a spring and damper in
series, the Mandel stress determines both the elastic strain and the
rate of plastic flow. Equations \ref{elastic_stress} and
\ref{flow_rule} are written in this parlance to give the
elasto-plastic constitutive relations:

\bigskip
\bigskip

\noindent \underline{Elasticity relation}:
\begin{equation}\label{final_elast}
\mb = 2G\mathbf{E^e_0}+\kappa \left(\text{tr}\mathbf{E^e}\right)\vec{1}
\end{equation}

\noindent where $\kappa=B\sqrt{\Delta}[1+\gamma^2/(2\Delta^2\xi)]$ and
$G=\sqrt{\Delta}B/\xi$, for $\Delta=-\text{tr}\vec{E}^e$ and
$\gamma=\sqrt{\vec{E}_0^e:\vec{E}_0^e}$. If $\text{tr}\mathbf{E^e}>0$,
both $\kappa$ and $G$ are 0.

\bigskip
\bigskip
\bigskip

\noindent \underline{Plastic flow rule}:
\begin{equation}\label{final_flow}
\vec{L}^p=\dpb = \frac{I_0}{d}\sqrt{\frac{P}{\rho_s}} \ \frac{\mu-\mu_s}{\mu_2-\mu}\frac{\mb_0}{\tau}
\end{equation}

\noindent where $P=-\text{tr}(\mb)/3$, $\tau=\sqrt{\mb_0:\mb_0/2}$, and $\mu=\tau/P$.  If $\mu< \mu_s$, then $\dpb=\mathbf{0}$.

\bigskip
\bigskip
\bigskip

As for the initial conditions, we declare that the granular body begins free of any plastic deformation, and that the body's initial deformation increment is elastic:
\[\fp(t=0)=\mathbf{1}, \ \ \ \ \vec{L}^p(t=0)=\vec{0}.\]
The granular model being proposed in this paper is fully defined by
the closed system of equations \ref{kroner}, \ref{Lp}, \ref{polar},
\ref{hencky}, \ref{balance}, \ref{mandel},
\ref{final_elast}, and \ref{final_flow} along with these initial
conditions.

\section{Implementation and results}\label{numerical}

The above equations were solved numerically using the ABAQUS/Explicit
software package (from ABAQUS 6.5) with the constitutive behavior
coded as a Vectorized User Material (VUMAT) model.  In line with our
interests, we choose three standard geometries that eventually have a
sustained, well-developed flow response:

\begin{itemize}
\item Long inclined chute
\item Annular Couette cell (with downward gravity)
\item Wide flat-bottom silo flow
\end{itemize}

The simulations are run long enough for transient behavior to
vanish. We may refer to this as ``steady'' flow, but a more general
meaning is implied, since environments like silo drainage do not
correspond to an Eulerian boundary value problem (the top surface
always descends). Beyond visual observation, the disappearance of
transients is quantifiable in the total kinetic energy, which becomes
constant (compared to the kinetic energy transients) when flow is
well-developed. 

\subsection{Numerical considerations: The method of inertial density reduction}

ABAQUS 6.5 implements Lagrangian finite element deformation,
presenting some challenges in the modeling of steady flow behavior.
In particular, there are two major numerical concerns that shall now
be addressed:

\begin{enumerate}
\item The steady response generally emerges after a high magnitude
  of deformation.  However, accuracy decreases dramatically when
  element distortion is too large.  Arbitrary Lagrange-Eulerian (ALE)
  routines can be used to counteract this by periodically sweeping the
  mesh to a less distorted configuration.  However, we find that the
  convection of variables between mesh sweeps carries unacceptably
  large error especially near boundaries, where spatial gradients have
  a reduced sample space.  Later versions of ABAQUS define purely
  Eulerian elements, which may improve this situation. However, as
  will be discussed momentarily, a different
  remedy can be instituted that uses the typical formulation. \\

\item The constitutive relations are increasingly sensitive in the low
  pressure regime.  By equation \ref{final_flow}, low pressure
  material at the onset of yield plastically flows at a rate
  proportional to $P^{-1/2}$, which can cause rapid variation in
  $\vec{D}^p$ during the transient phase.  From an implementation
  standpoint this induces numerical stiffness; the rate of change may
  be too large to be adequately represented by the stable time
  increment.  Furthermore, by equation \ref{final_elast}, the elastic
  moduli vanish sharply with decreasing pressure: $\kappa\propto
  P^{1/3}$ as $P\rightarrow 0$.  Once again, this requires that the
  time step be small enough at low pressures so as to accurately track
  the changes in elastic properties.  Numerical stiffness of the type
  just described is an important consideration since several of the
  flow environments to be investigated utilize the traction-free
  boundary condition, inducing non-negligible regions of low pressure
  in the material body.

\end{enumerate}

A single encomassing remedy, novel to this author's knowledge, was
instituted to resolve both these issues: the inertial density of the
material was artificially reduced by several orders of magnitude,
holding the gravitational density fixed.  The true granular material
density (roughly $1500$ kg/m$^3$ for glass beads) was used in body
force computations in order to accurately represent the force of
gravity. However, the density used in computing the inertial force
$\rho \dot{\vec{v}}$ was artificially decreased by factors on the
order of $10^4-10^5$.  In systems without excessive curvature, the
inertial force should vanish in steady state, so this alteration does
not impact the correctness of the well-developed flow solution.  To be
clear, density reduction is performed only on the unstressed reference
body, so that mass conservation is upheld during flow.

By decreasing the inertial density, the wave speed increases, causing
transients to pass quickly in physical time.  This allows the steady
state velocity and stress fields to emerge faster than the time
necessary for large distortions to occur to the body (and likewise the
mesh). A clear analogy for this effect can be observed in a
rudimentary mass/spring/damper system in series--- under applied
force, the distance over which the motion of the mass is unsteady
decreases as mass decreases. Using this technique, most simulations of
the granular model reached steady state before any nodal displacements
were noticeable to the eye. At the same time, since the stable time
increment is proportional to $\sqrt{\rho}$, high enough density
reduction reduces the time step (relative to the deformation speed)
below the threshold necessary to accurately represent low pressure
material during the transient phase. Hence, artificial density
reduction confronts both issues raised above. It should be noted
though, that there is no apparent benefit in terms of
\emph{computation time} when using this method since the transient
time period and the stable time increment both decrease. More details
on the method can be found in \cite{kamrin_thesis}.

\subsection{Verifying the numerical model}\label{test}

Throughout this work, the model's six parameters are assigned the
following values:

\begin{center}
\begin{tabular}{|l|l|}
\hline
$B=7\times 10^9$ Pa & $I_0=0.279$
\\
\hline
$\rho_s=2450$ kg/m$^3$  &  $d=0.003$ m
\\
\hline
$\mu_s=\tan(20.9^{\circ})$ & $\mu_2=\tan(32.76^{\circ})$
\\
\hline
\end{tabular}
\end{center}
Recall that $\xi$ is also a parameter, but its value is tied directly
to the value $\mu_2$. Except for $d$, these values were all lifted
from Jiang and Liu \cite{jiang03} and Jop $\etal$ \cite{jop06}.  As
both groups considered spherical glass beads, it is assumed that their
data should be representative of the same material.  The particle
diameter is set to $3$mm, as this is the common value used in the
experiments and DEM simulations of the MIT Dry Fluids Group, whose
data this model will be compared against later. It should be noted
that the particle diameter easily scales out in the
non-dimensionalization.  The elements used, unless otherwise stated,
are hexahedral of type C3D8R in ABAQUS/Explicit.


To check the implementation of the material model, one element tests
were performed wherein a box element is compressed laterally with a fixed
pressure, and sheared tangentially with various shear
tractions. Implicit and explicit numerical integration routines were
encoded into VUMAT files, and it was verified that both give
steady flow rates that match the analytical form (equation
\ref{rheology}) to high accuracy.
Since explicit integration is only conditionally stable, the time step
must be sufficiently small in order for numerical stability at the
higher shear stress values.  The implicit routine is always stable
numerically, but for higher shear stresses, the time step must be
reduced to ensure convergence of the Newton-Raphson solver.  For the
simulations detailed in this work, the time step is always small
enough for explicit stability, a byproduct of the artificial density
reduction.  Together with the fact that explicit integration is
computationally faster, the choice was made in the upcoming
simulations to use the explicit routine.

\subsection{Rough-walled inclined chute}

\begin{figure}[t]
(a) \epsfig{file=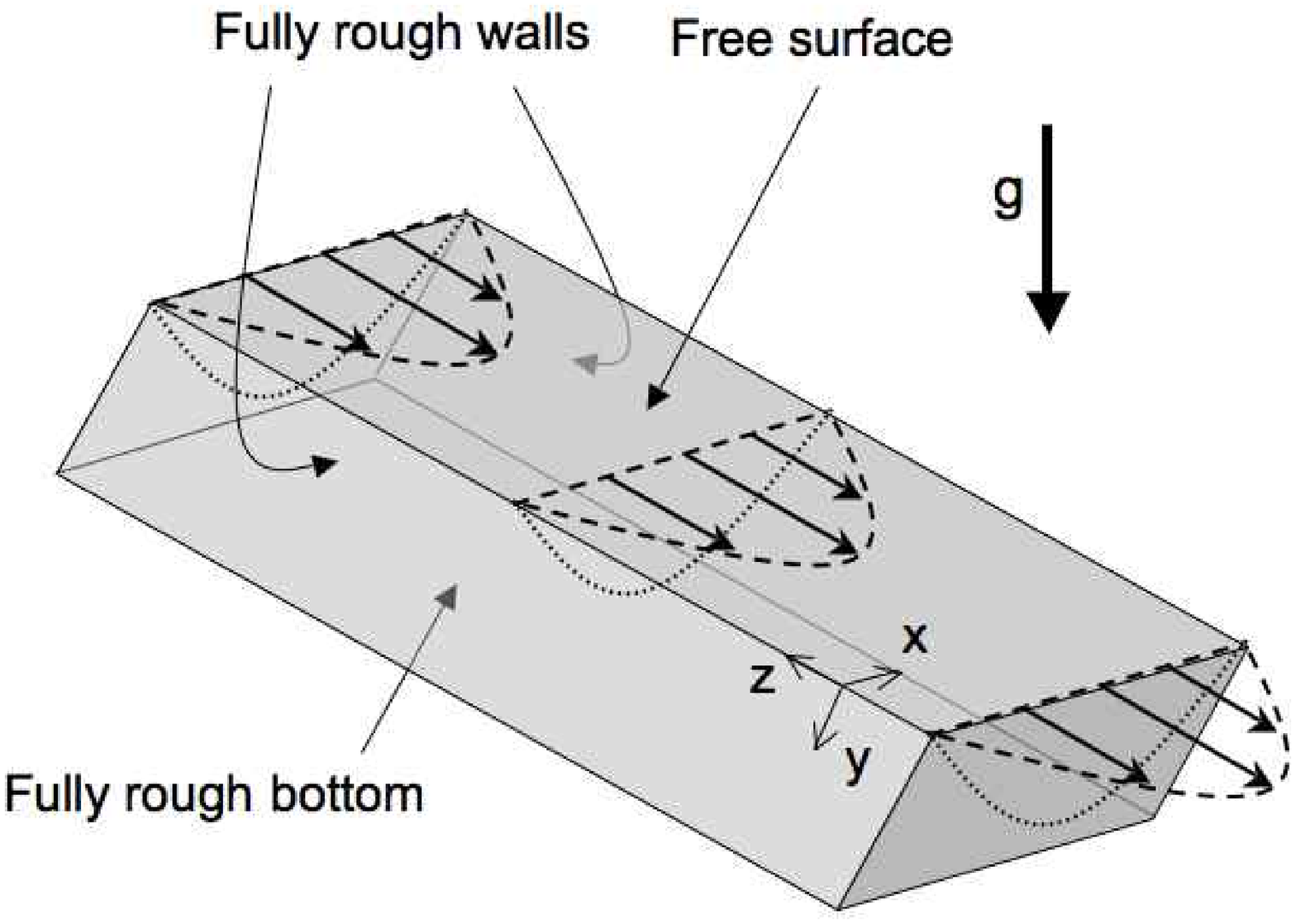, width=2.3in, clip} \ \ \ \ \ \ (c) \ \ \ \epsfig{file=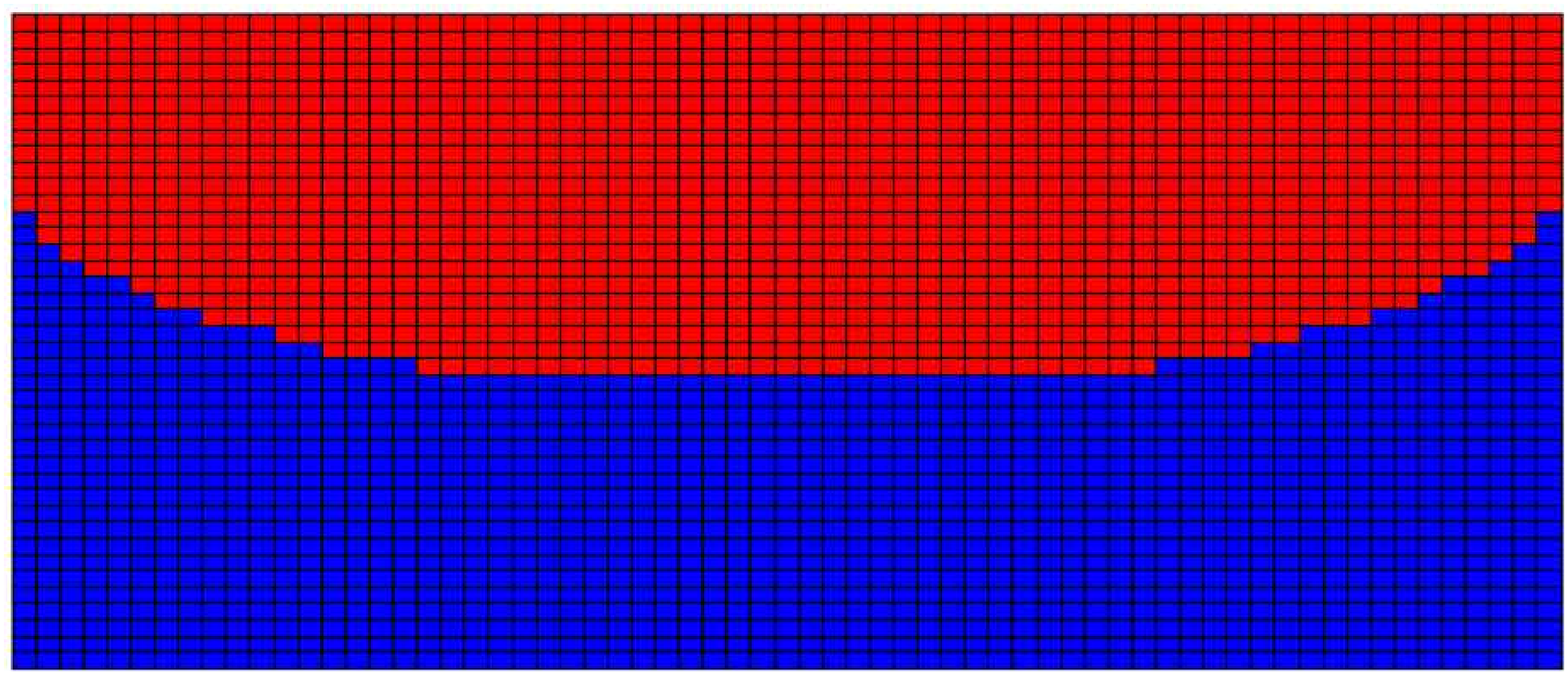, width=2.7in, clip}

\bigskip
\bigskip
(b) \epsfig{file=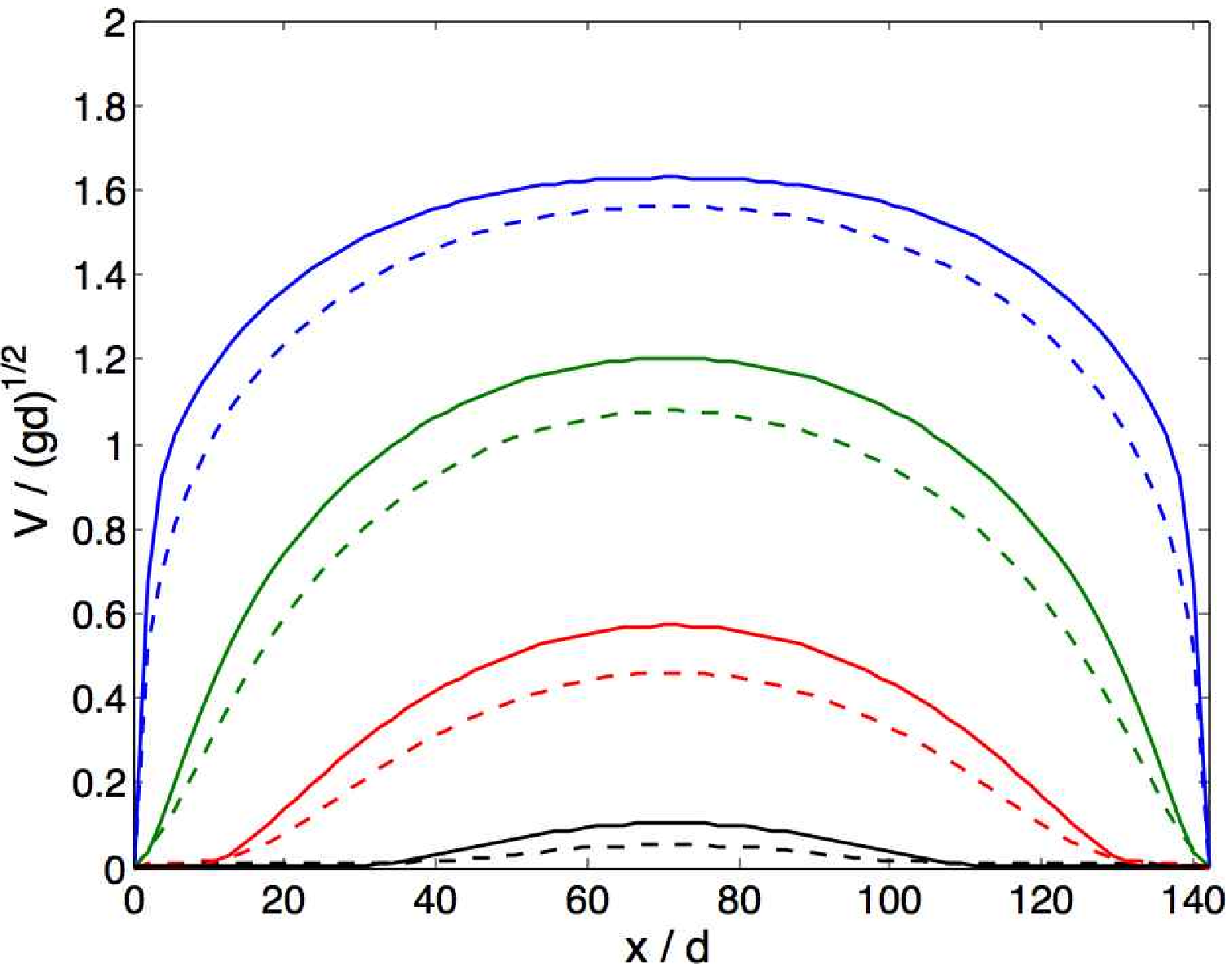,width=1.9in,clip} \ \ \ \  (d)\epsfig{file=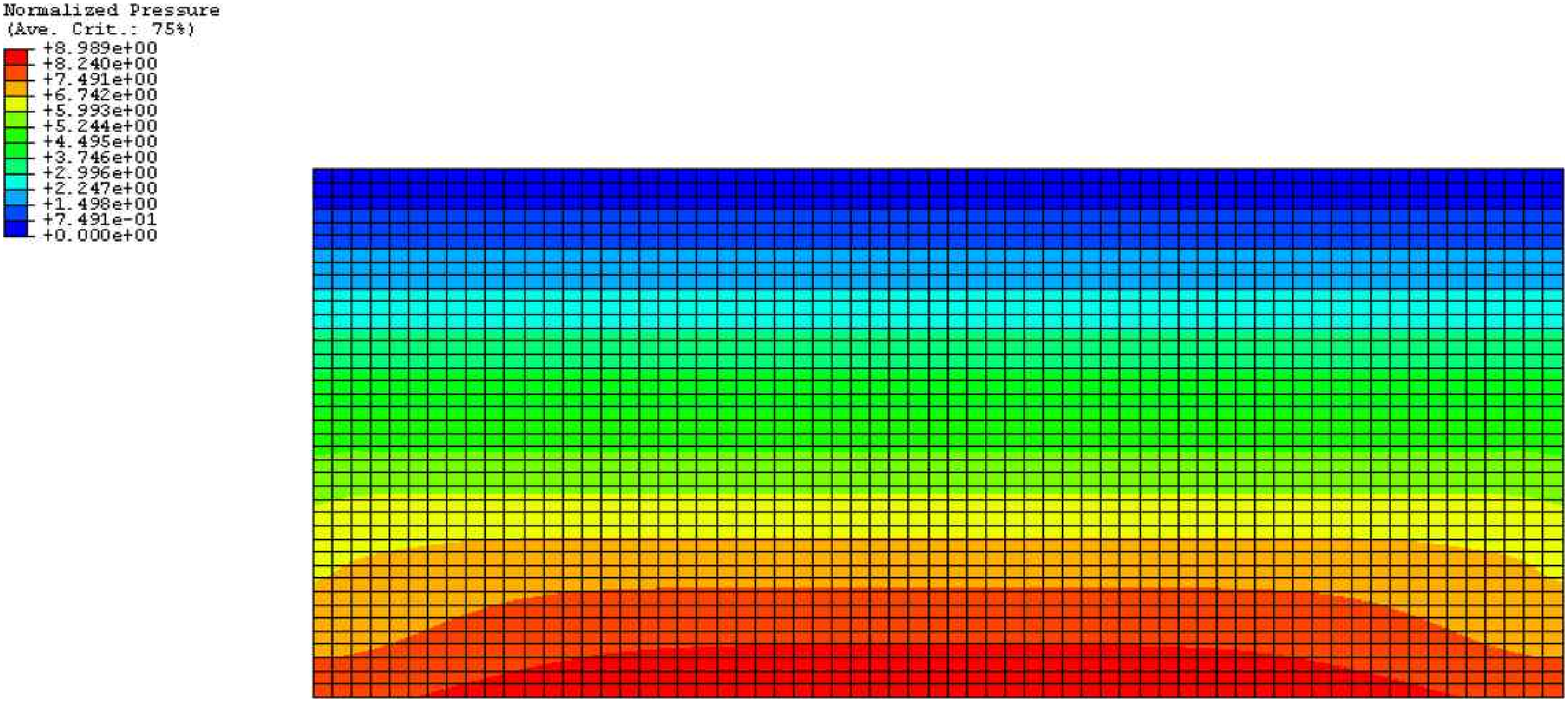,width=3.4in, clip}
\caption{(a) The rough inclined chute setup. (b) The velocity as a
  function of $x$ at depths $y/d=0,9.2,18.4,27.6$. The dashed curves
  are the numerical results of Jop $\etal$ \cite{jop06} via finite
  differences with a Stokes-type solver.  The solid curves are from
  the steady solution of the elasto-plastic model implemented on
  ABAQUS/Explicit. (c)-(d) Elasto-plastic results viewing the $xy$ plane with $y$ downward:  In (c), regions of plastic flow are in red and static in blue. In (d), the normalized pressure $P/\rho g d$ is displayed.
}\label{incline_results}
\end{figure}

Figure \ref{incline_results}(a) reviews the geometry and boundary
conditions for the rough inclined chute environment, which was
originally studied by Jop $\etal$. In that work, equation
\ref{flow_rule} was directly solved as a non-Newtonian Bingham fluid,
letting viscosity go to $\infty$ below yield. The approach delivers a
steady velocity computation (with some numerical difficulty near the
rigid interface), but by avoiding elasticity altogether, no stresses
can be computed in the rigid zones. In terms of completeness, the
Bingham fluid approach is useful, but not equipped to handle the
general traction/kinematic boundary value problem. The elasto-plastic
formulation herein is complete but has added many new relationships
and a different kinematical perspective on deformation. However, the
predicted steady flow should not differ noticeably from the Bingham
flow, as elastic stretching vanishes in steady-state. By analyzing
this case, we hope to show sufficient agreement between the two
solutions, ensuring the elasto-plastic model has inherited the
successes of the Bingham flow solution in this environment
\cite{jop06}.

The chute is long such that the flow can be deemed invariant in the
$z$ direction.  The bottom of the chute is at $y=60d$ and the chute is
$142d$ wide. The angle of incline of the chute with respect to the
horizontal is $22.6^{\circ}$.

The chute model uses $2600$ box elements arranged in a grid $65$
elements wide in the $x$ direction by $40$ in the $y$. The assembly is
only $1$ element wide in the $z$ direction--- symmetry conditions were
invoked by constraining the nodes on the back face of the $xy$ slab to
move with the same displacements as their front face counterparts. The
simulation must start in a reference state with no gravity.  Gravity
is then ramped up to its full value.  Since the material must be
compressed before any shear stress can be supported, the components of
gravity are applied one at a time; the $y$ body force is first
smoothly ramped up over $1\times 10^{-5}$s to its final value of
$f_g\cos(22.6^{\circ})$, where $f_g=0.63 g\rho_s$. After a $2\times
10^{-6}$s pause for any needed relaxation, the component pointing down
the chute is smoothly ramped up to $f_g \sin(22.6^{\circ})$ over
$2\times 10^{-5}$s.  The assembly is then left to flow until a total
time of $5\times 10^{-4}$s has elapsed.

The inertial density was artificially decreased by a factor of several
hundred thousand to a value of $\rho=2\times 10^{-3}$. At the free
surface, theoretically, the pressure goes to zero causing the elastic
moduli to vanish. This is dangerous for procedures such as
ABAQUS/Explicit that consistently send waves through the material---
any small perturbation could cause a node to accelerate out of
control.  Moreover, the flow rule becomes undefined.  To keep the free
surface in tact, a few Pascals of overpressure are applied.

While the flow is eventually very steady to the eye, a more
quantitative measure is desired. From $t=0-4\times 10^{-5}$s, the
motion is markedly transient due to gravity ramping up.  The system
then relaxes toward the steady state, first rapidly but then slowly as
steady state is approached. Comparing the system's rate of relative
kinetic energy change in the fast relaxation period to its value at
the end gives a rough criterion for how steady a flow is.  At the end
of the run, the rate of decrease of the relative kinetic energy is
over $500$ times smaller than the value it takes during the initial
relaxation, which indicates a strong steady-state behavior.

Comparisons to the numerical results of Jop $\etal$ are displayed in
figure \ref{incline_results}(b).  The agreement is quite good considering
how disparate the solving methods are. Jop utilized a fixed grid
finite-difference scheme solving a non-Newtonian Navier-Stokes-type
problem, while the elasto-plastic results were obtained by a
Lagrangian explicit procedure on nodes that are constantly moving
during steady-state.  While differences of numerical procedure can
cause different sources of gain and loss, it seems more likely that
the discrepancies are stemming from the free surface condition.  The
true deformation rate at the top corners is actually infinite.  As
this is numerically impossible for either scheme, large but finite gradients
occur there numerically as determined by the choice of the free
surface treatment. This has a clear trickle-down effect on the global
velocity field and, in lieu of the particular fashion in which the
solutions differ, we suspect this effect causes much of the
difference between the two numerical data sets.

One distinguishing feature of the elasto-plastic model is its ability
to calculate both flow and stress everywhere. Figure
\ref{incline_results}(d) displays the pressure distribution over the
full geometry. Note that the stresses vary smoothly through the
transition from yielding to static, as opposed to the interfacial
stress issues that commonly occur with Bingham fluid models. In
particular, observe that the pressure field goes from decreasing
linearly in the flowing zone, to decreasing somewhat nonlinearly in
the static zone.  In this geometry, compressive stresses in the $x$,
$y$ and $z$ directions of the steady flowing zone must all be
identical under codirectionality. A hydrostatic pressure profile is
induced as a result. But upon descending below the flowing layer,
codirectionality no longer has this influence on the stresses and a
somewhat more complicated elasto-static form for the pressure field
becomes apparent.  Discrete simulation data for this environment would
be helpful for checking the validity of the stress profile in the
static zone.  Notably, the primitive fully-rough boundary conditions
used may not be the most accurate reflection of the true conditions on
the bottom and side walls.

\subsection{Annular Couette cell}

\begin{figure}[t]
(a) \epsfig{file=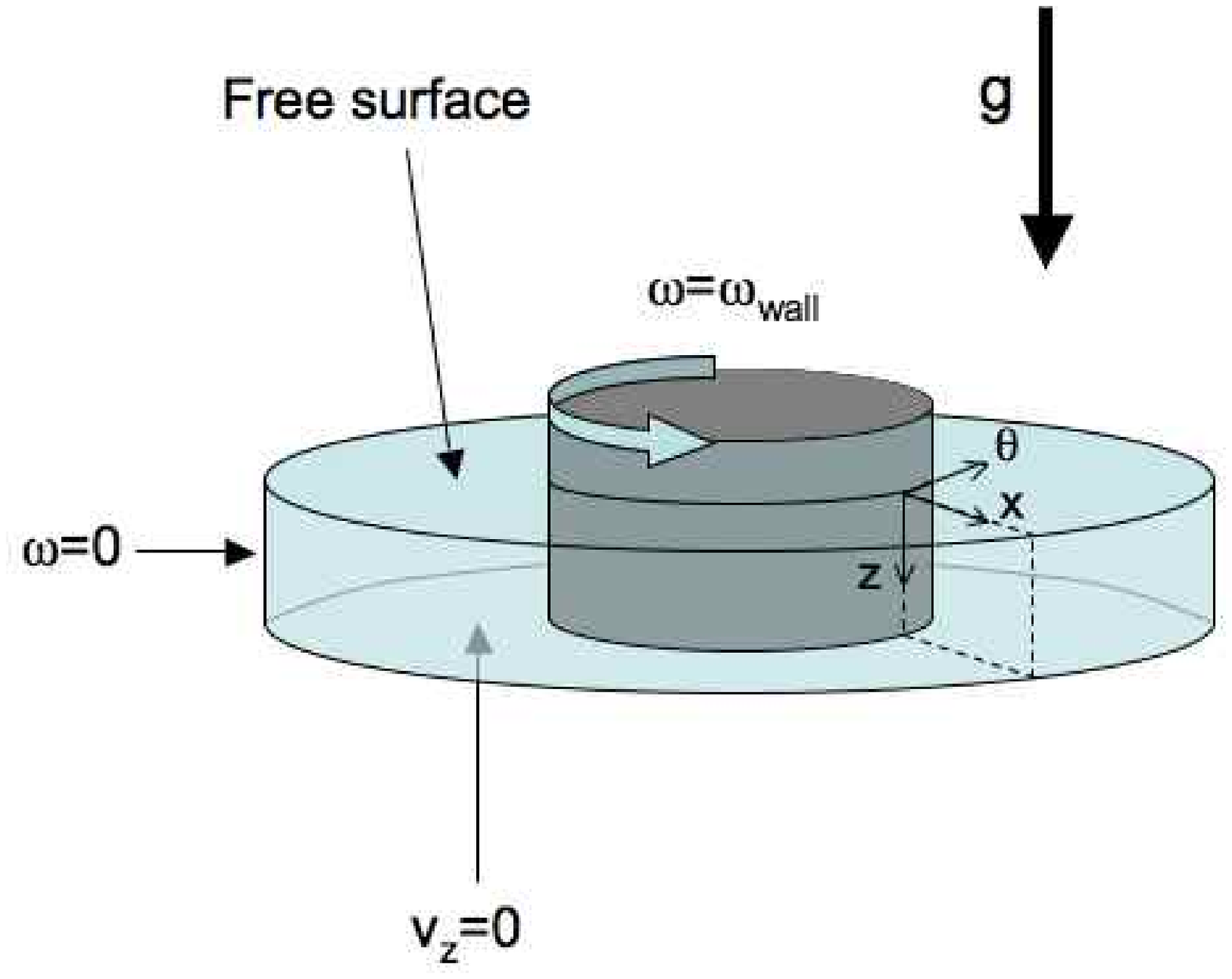, width=2in, clip} \ \ \ (c)\epsfig{file=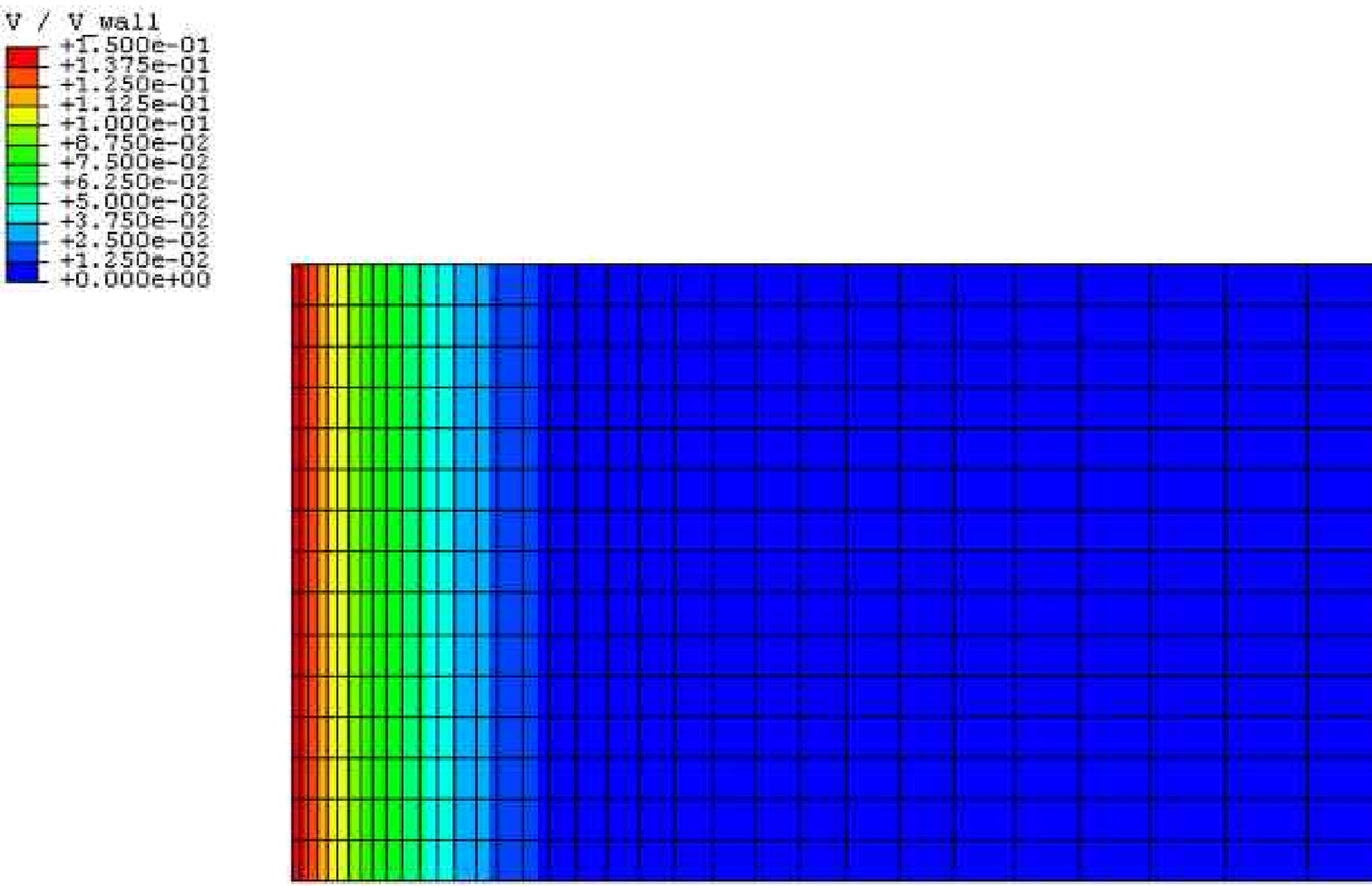, width=3.5in, clip}

\bigskip
\bigskip

(b)\epsfig{file=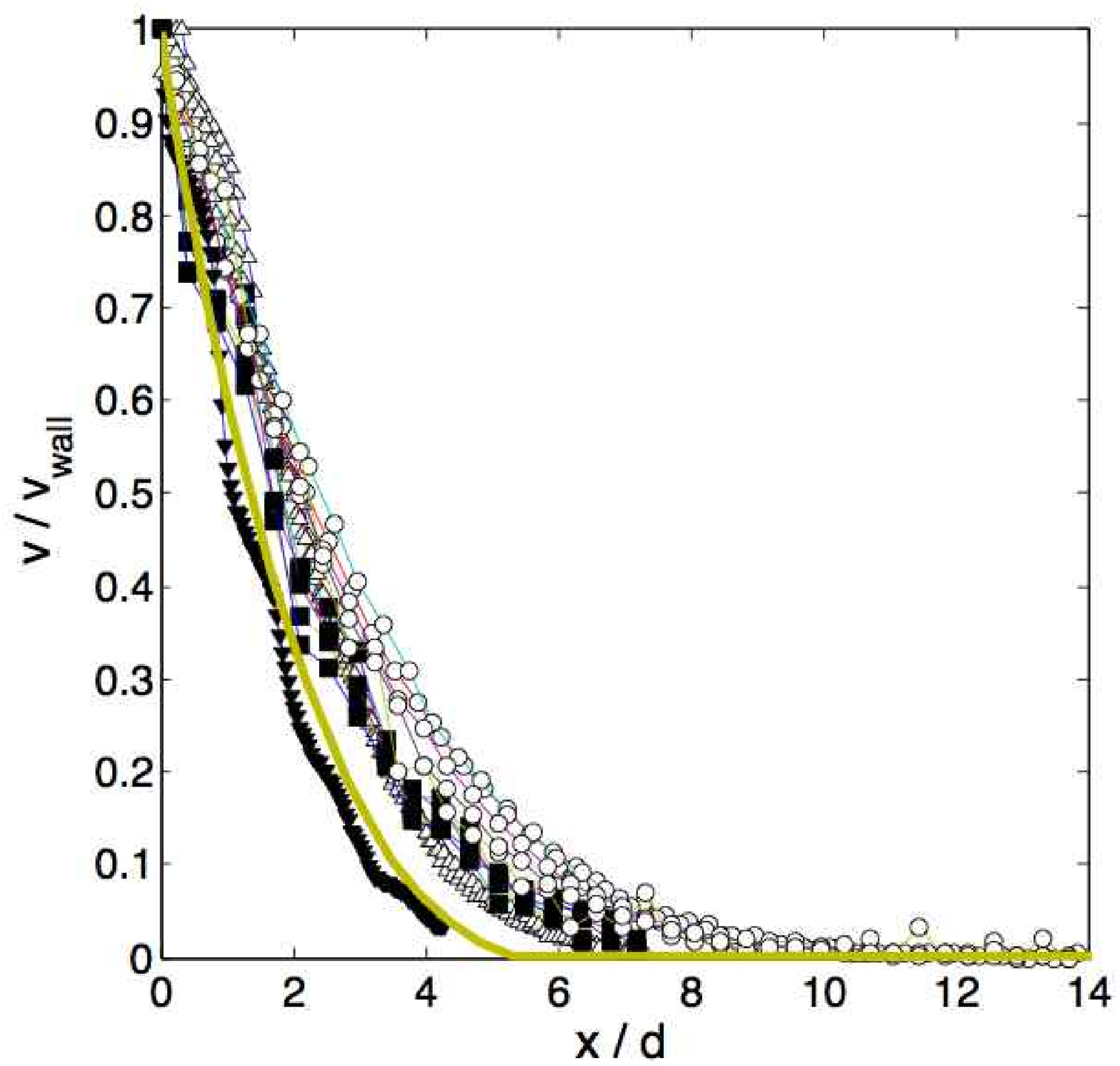, width=2in, clip} \ \ \ (d)\epsfig{file=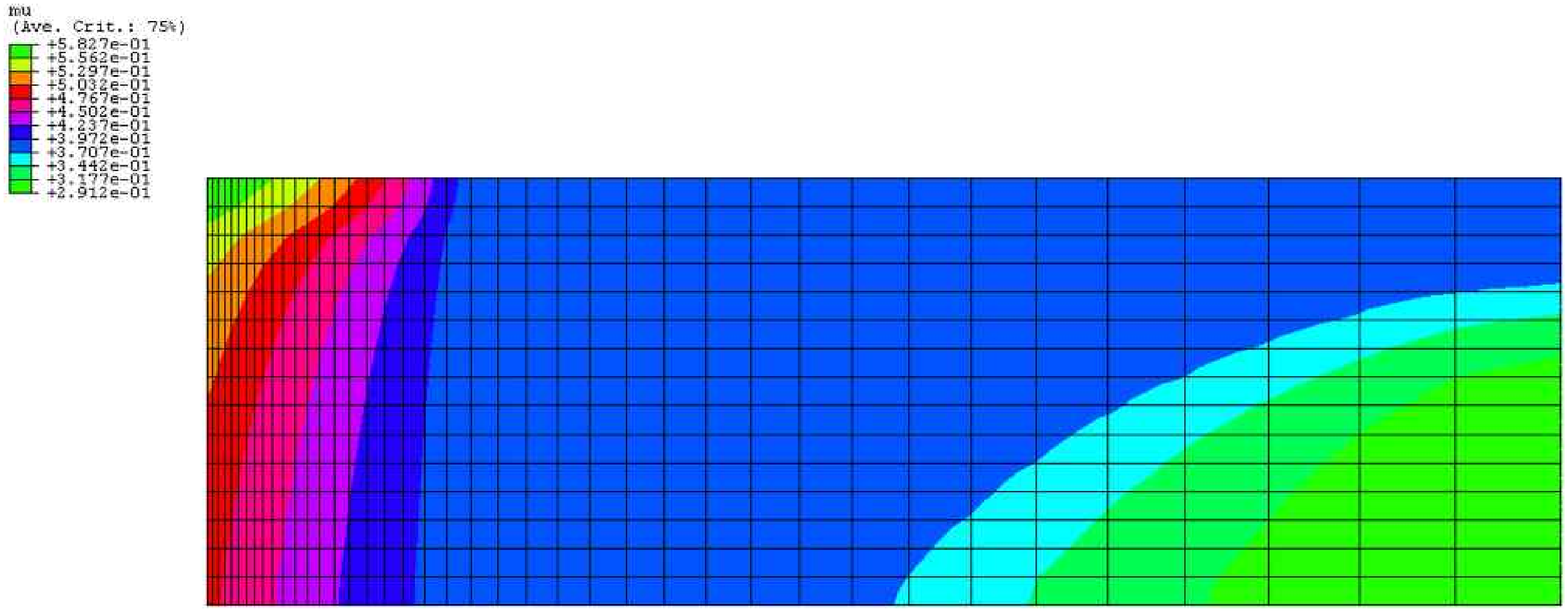, width=3.5in, clip}
\caption{(a) The annular Couette setup. (b) The velocity profile
  normalized by the wall speed as predicted by the elasto-plastic
  model at half-height (thick green line) compared against fifteen
  experimental and discrete element simulation data sets for this type
  of flow as compiled in \cite{midi04}. In (c)-(d) the $xz$-plane is
  shown with $z$ downward: (c) The velocity normalized by the wall
  speed. (d) The stress ratio $\mu$.}\label{annular_results}
\end{figure}

We now move on to the annular Couette cell, an environment which has no
previously known solution to the Jop \etal flow law. The
results shall be compared directly against the myriad of data on this
environment compiled by G. D. R. Midi \cite{midi04} and thus the
geometric specifics and boundary conditions were chosen so as to give
a good representation of the conditions used in these studies.
Referring to figure \ref{annular_results}(a) for general details on the
environment, the following values were selected:
$\omega_{\text{wall}}=1.25$rad/s $\approx 0.2$rev/s, the distance from
inner to outer wall is $x_{\text{out}}=30d$, the height is
$z_{\text{bottom}}=10d$, and the inner wall radius is $40d$.  At the
walls, the material motion must match the wall motion in the $\theta$
and $x$ directions, but material can slide without resistance up and
down the walls.

Since flow and stress should be symmetric in the $\theta$
direction, the behavior as seen in a downward cut through the annular
trough should represent the global behavior. A narrow sector of the
annulus (total angle $0.1^{\circ}$) was likewise simulated using
periodic boundary conditions on the front and back faces--- nodal
displacements on the front face are constrained to be identical to
those on the back face except rotated appropriately by
$0.1^{\circ}$. The sector is modeled using $40$ elements in the $x$
direction, $15$ in the $z$, and a thickness of one element in the
$\theta$ direction, for a grand total of 600 elements.  Almost all the
motion is known to occur near the inner wall in this
environment. After preliminary tests of the elasto-plastic model
produced the same conclusion, a bias was utilized to crowd nodes
closer to the inner wall and improve precision.  The bias resulted in
half of the elements occurring within a distance of $6d$ from the inner
wall.

The inertial density was reduced by a factor of over one hundred
thousand to a value of $\rho=0.01$. First, gravity is smoothly ramped
up to its full value over $t=0-1\times 10^{-5}$s.  During this time
period, as before, a slight overpressure is also applied to the free
surface for stability. The rotation of the inner wall then commences,
smoothly ramping up to a final value of $\omega_{wall}=1.25$rad/s over
$t=1-2.5\times 10^{-5}$s. The simulation is then left to flow until a
total time of $t=1\times 5^{-4}$s has elapsed.

Once the inner wall rotation reaches $\omega_{wall}$, the flow soon
after begins to relax toward a macroscopically visible
steady-state. Evidencing the fact that the system has indeed reached
sufficient steady-state by the end of the run, the rate of kinetic
energy loss of the system at the end of the simulation is over three
orders smaller than the loss rate immediately after the ramp-up.


Observing figures \ref{annular_results}(b) and
\ref{annular_results}(c), we notice a few major qualitative
points. For one, the flow forms a clear shear band near the inner
cylinder, a fact corroborated by the provided data.  While the
predicted thickness is somewhat smaller than found experimentally, the
fact that it is on the same order is a major highlight for a model with
no added fitting parameters (though we take this point with a grain of
caution as shall be discussed momentarily). The authors of
\cite{jop06} expressed their belief that the flow rule would be
incapable of describing narrow shear bands.  However, the above
results show quite clearly that this is not the case for the
elasto-plastic model.

Another observation is that the velocity profile does not vary to any
observable extent in the $z$ direction.  This result has been verified
in DEM simulations of this environment \cite{kamrin07b}, where it was
found that almost no vertical fluctuation occurs in the fast zone near
the inner wall. Of course, the elasto-plastic velocity field dies off
much differently than experiment.  Where the experimental data is
shown to be well-fit with an exponential decay that extends throughout
the Couette cell, the elasto-plastic solution predicts a sharp cutoff
around $5.5d$ from the inner wall. This result should come as no
surprise, since the model does not account for quasi-static, non-local
behavior, of which slow exponential decay is a textbook case.

Viewing figure \ref{annular_results}(d), it is clear that while the
velocity appears invariant in the $z$ direction, other important
stress-based quantities are not. The stress ratio $\mu=\tau/P$ indeed
has a highly non-trivial spatial distribution, reflective of the
nonlinear dependence of the stress state on the plastic flow.

Experimental data on this flow environment indicate that the shear
band width stays relatively constant for a range of slower wall speeds
\cite{midi04}, and grows with increasing wall velocity in the faster
flow regime \cite{koval_thesis}.  As expected from rate-sensitivity,
we have verified with multiple simulations that the elasto-plastic
model agrees with the latter case: the shear band width correlates
positively with inner wall speed (holding gravity constant). We expect
this correlation to falsely persist even as the wall speed becomes small.
This point serves as a careful reminder that the model has no
internal length-scales and requires user discretion when assessing
flow predictions--- one must check that the normalized flow rate and
shear-rate gradients obey the limitations outlined in sections
\ref{regimes1} and \ref{continuum}.

\subsection{Flat-bottomed silo}
While the past two environments have focused on comparing velocity
profiles, the stresses have yet to be directly tested.
Experimentally, the stress tensor is a difficult quantity to measure.
The stresses in 2D disk assemblies can be approximated using
photoelastic grain material, however, there is not currently an
experimental method available to measure the stress tensor within an
arbitrarily flowing 3D granular material.  For this measurement, the
best option as yet is to utilize DEM and compute the local stress
tensor from grain fluctuations and inter-grain contact forces per
equation \ref{discrete_stress}.

\begin{figure}[t]
\begin{center}
(a) \epsfig{file=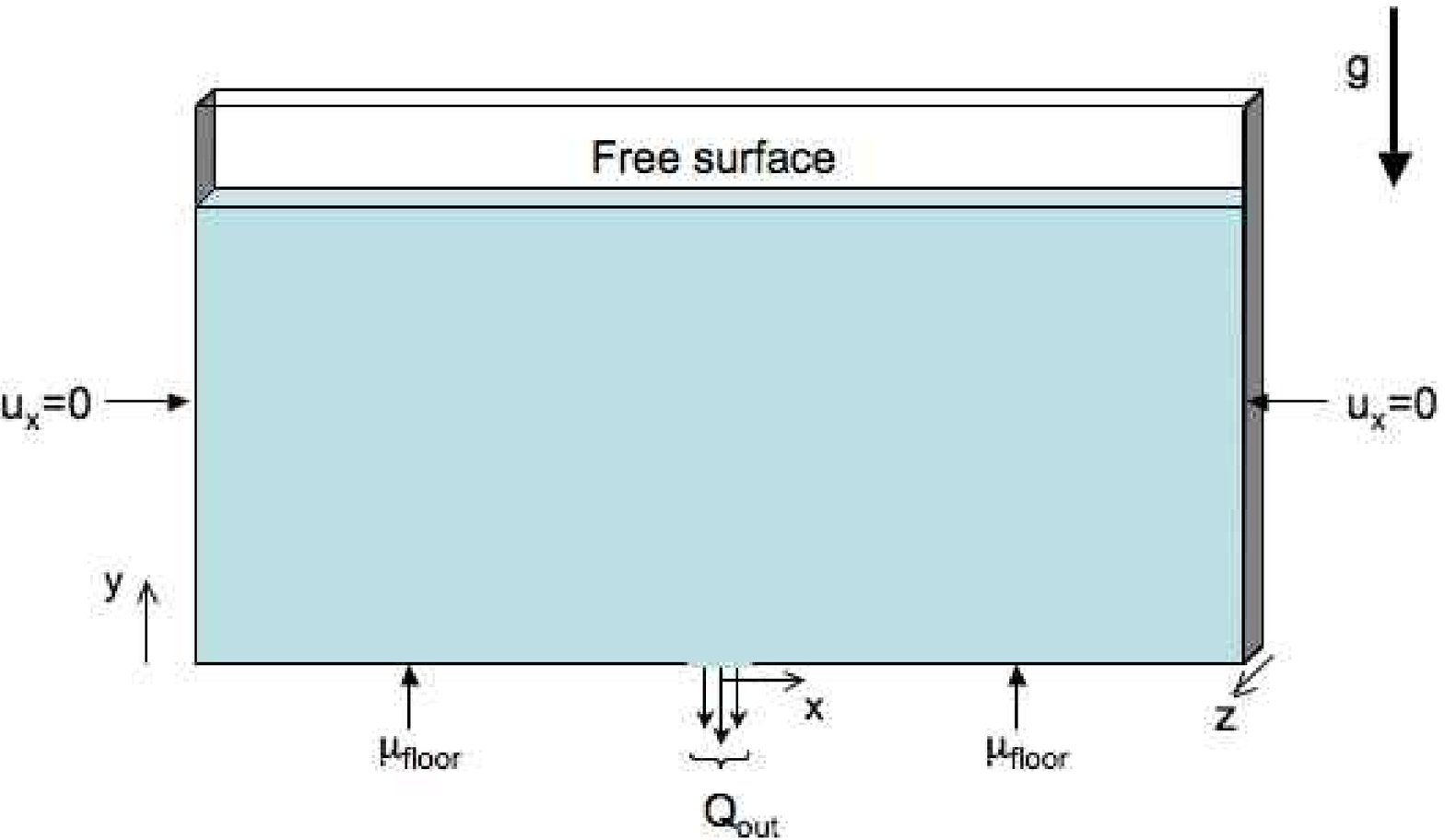, width=2.9in, clip} \ \  (b)  \epsfig{file=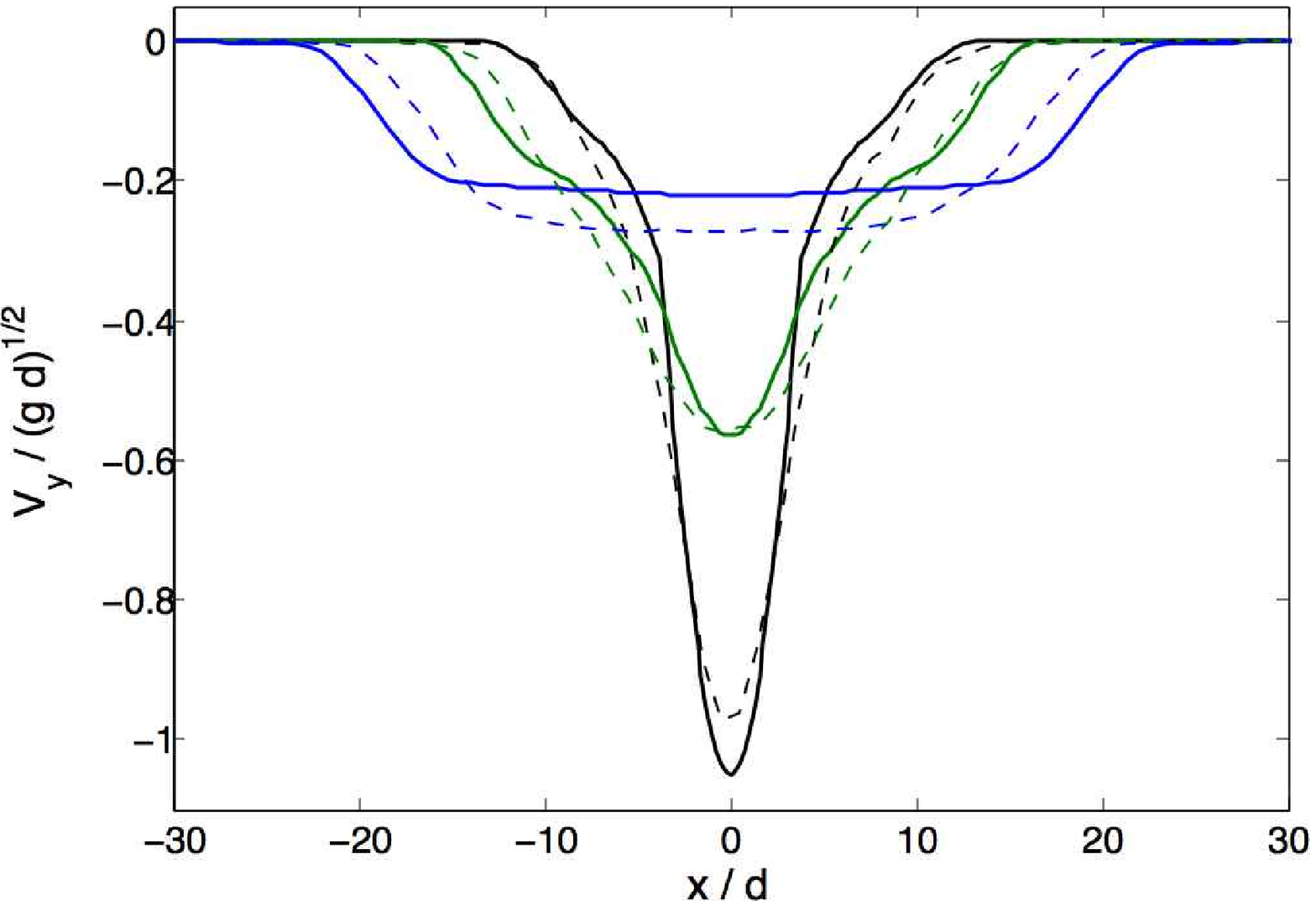, width=2.3in, clip}

\bigskip
\bigskip

(c) \ \ \ \ \epsfig{file=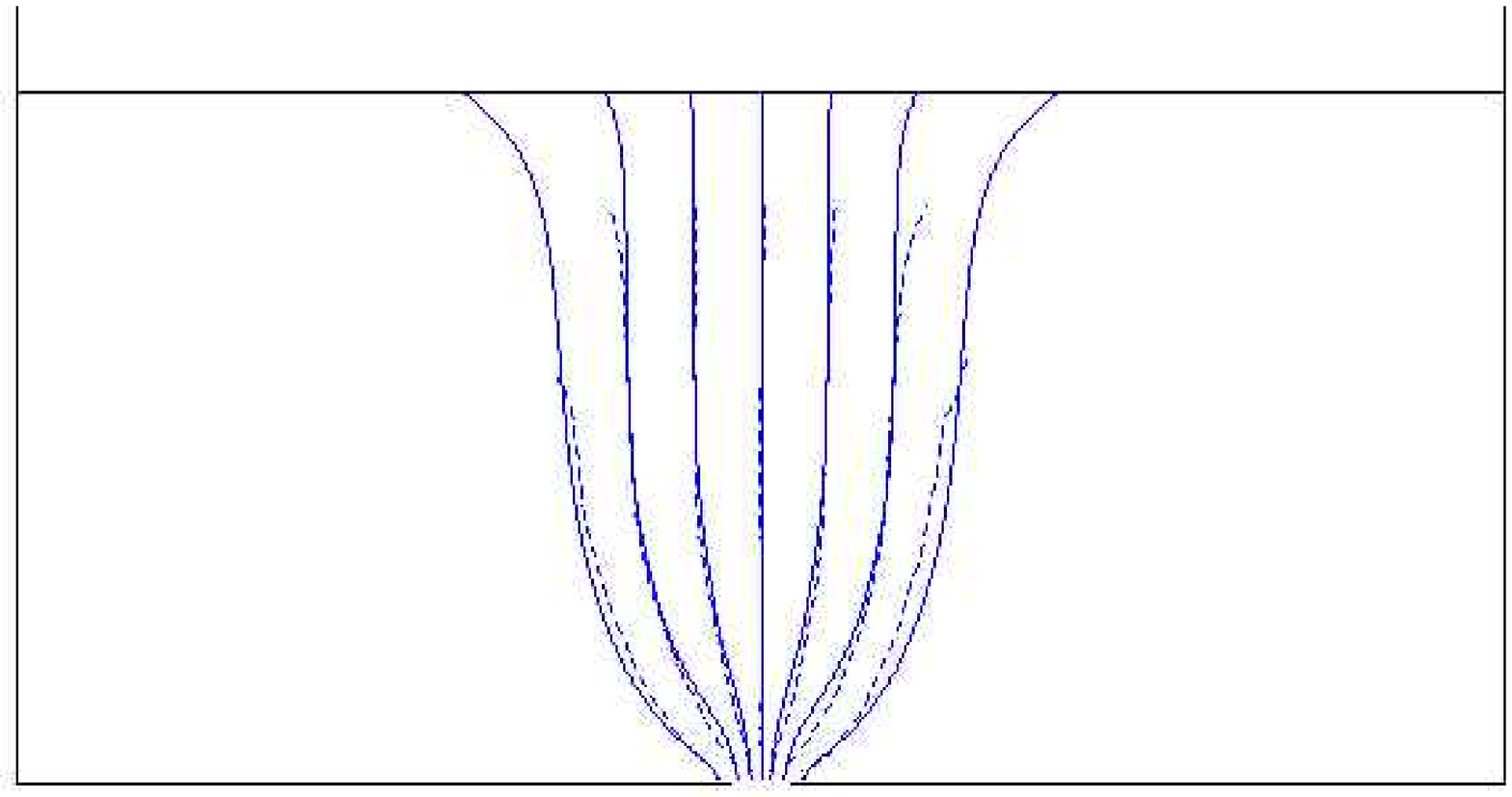, width=3.3in,clip}
\caption{(a) The flat-bottomed silo setup. In (b)-(c), a comparison of
  elasto-plastic velocity results (---) to DEM data (- -) c/o
  Rycroft \cite{rycroft09}. (b) The $y$ velocity component as a
  function of $x$ at heights $y=5d, 10d, 30d$. (c) Trajectories
  predicted by the elasto-plastic model alongside the DEM
  trajectories. (Container outline provided for ease of
  viewing.)}\label{silo_vel}
\end{center}
\end{figure}

Rycroft \etal \cite{rycroft09} have performed DEM simulations of wide
silo drainage.  The computed stress and flow fields shall now be
compared directly against the predictions of the elasto-plastic
model. A schematic diagram of the flow geometry is pictured in figure
\ref{silo_vel}(a). In accordance with Rycroft's simulations, we model
the silo as having an opening width of $6d$, a height of $70d$, and a
full width of $150d$. Rycroft enforced periodic boundary conditions in
the $z$ direction giving the silo an apparent $z$ thickness of $8d$
but without wall-ordering effects.  Since the flow should not
fluctuate in the $z$ direction, we simply enforce plane-strain
conditions (using plane-strain elements of type CPE4R).  Furthermore,
the silo has left-right symmetry about the vertical center-line, which
we take advantage of by modeling only the right half of the silo.

The floor of the silo is modeled as having a frictional interaction
with the material characterized by a coefficient of friction
$\mu_{\text{floor}}=0.2$.  This number was estimated loosely from
Rycroft's simulation--- to account for the interaction induced by the
effects of particle rolling, rearranging, and dragging along the
floor, the element/floor interaction was modeled by reducing Rycroft's
floor/particle friction coefficient by $60\%$.  Future study would be
necessary to determine the complete and precise form for element/floor
sliding interactions. While Rycroft's simulation utilized side walls
made of the same frictional material as the floor, for simplicity, the
elasto-plastic simulation employs the simpler condition of no $x$
displacement at the side-walls.  This could be enhanced in the future,
but in wide silos, the details of the side-walls have only a small
effect on the dominating behavior.

Ideally speaking, the boundary condition at the silo orifice should be
zero stress tractions.  However, this is highly problematic
numerically. In reality, silo flows develop a ``free-fall arch''
directly above the orifice \cite{nedderman}. The arch is typically
only a few particles high and connects the edges of the opening. Once
a particle passes through this hypothetical arch, it enters free-fall
and becomes gas-like.  A granular material element within the
free-fall arch would realistically have a smaller packing fraction,
but still support compressive stresses through internal random
particle collisions.  The elasto-plastic model, however, does not
include gas-like effects. Such dilation would be assigned to the
elastic deformation, causing the elastic moduli to vanish.

Our interest is not in the details within the immediate vicinity of
the orifice, rather the bulk material behavior within the greater silo
apparatus. However, with a zero-traction condition at the opening, the
situation described above destabilizes the simulation
prematurely. While free surfaces are usually taken care of by adding a
slight amount of artificial compression, this remedy will not suffice
here because the applied pressure has too much of an effect
on the evolution of the outflow rate.  

We are left with the alternative of using kinematic boundary
conditions at the orifice.  It would be overreaching to assign any
particular velocity profile at the orifice. Instead, we fix the total
flux out the orifice and let the material response determine the shape
of the flow profile.  To match the outflux in Rycroft's simulation,
\[Q_{\text{out}}/2=\int_{\text{Right
    half-opening}}v_y(x,y=0)dx=2.19\times 10^{-3} \
\frac{\text{m}^2}{\text{s}}\] was instituted at the orifice, encoded
as an equation constraint in ABAQUS.

Not far from the opening, large inhomogeneous deformation occurs at
small length-scales, necessitating many small elements to maintain
accuracy. A grand total of 9750 elements were used in modeling the
half-silo.  To minimize discretization error, the orifice was modeled
with a half-width of 15 elements. The adjacent silo floor was modeled
60 elements wide. The silo height was modeled with 130 elements. The
element width was constant within the orifice, but bias was used along
the other boundaries to maintain smooth changes in element sizes
throughout.  Elements shrink vertically as a sole function of distance
from the silo bottom. The element width is uniform at the top surface,
but moving downward, elements crowd the center so the floor region has
a smooth transition from wider elements near the wall to narrower
elements adjacent to the orifice.

Due to the high number of elements and small minimal element size to
system size ratio, this flow is too computationally intensive for one
one processor.  Instead, the 12-node Truesdell cluster of the MIT
Solid Mechanics Group was employed to solve the problem in parallel.
Using domain-level parallelization, the cluster split the half-silo
into 12 spatial regions.

\begin{figure}
(a)\epsfig{file=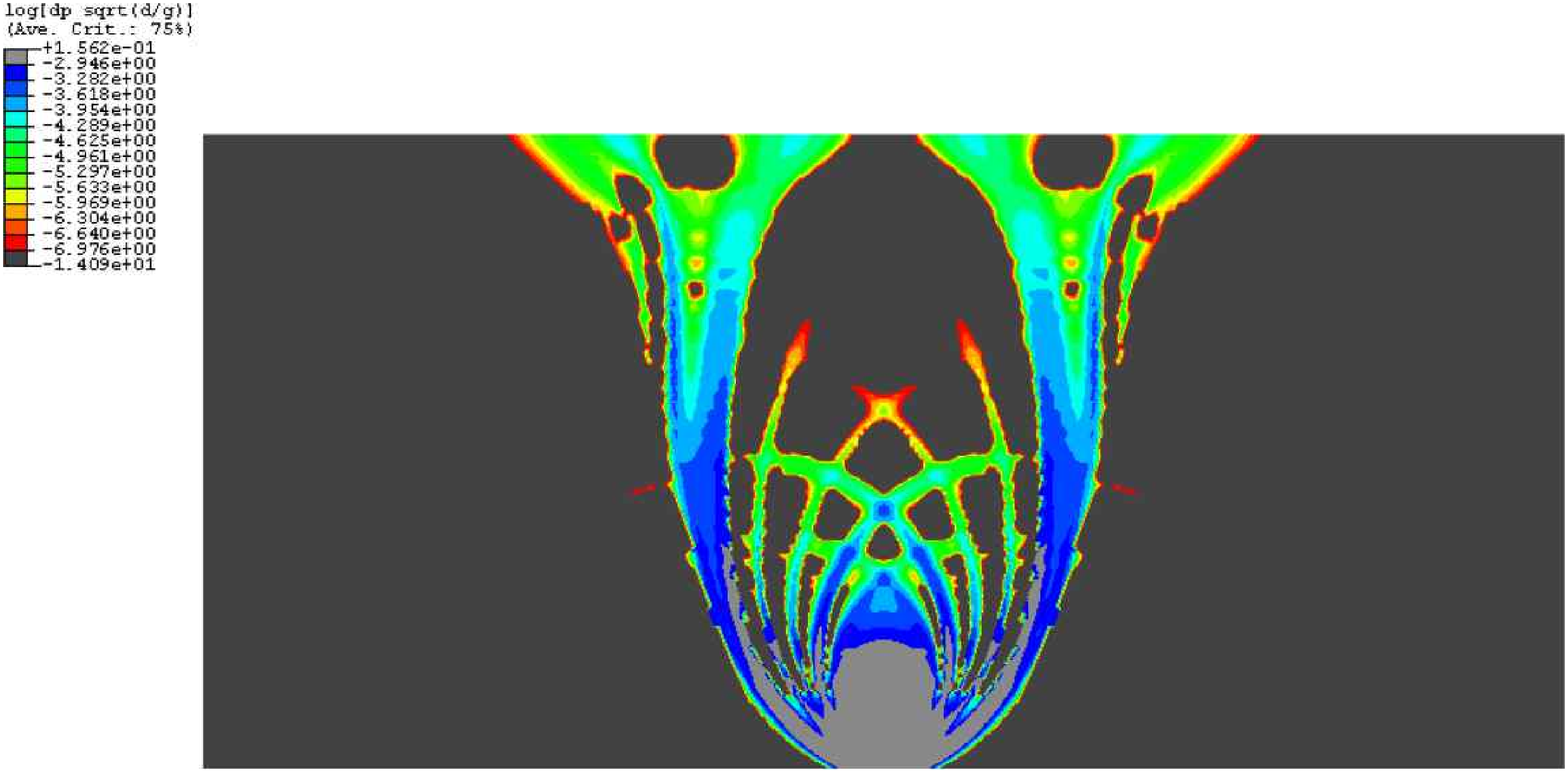, width=5.7in ,clip}

(b)\hspace{.64in}\epsfig{file=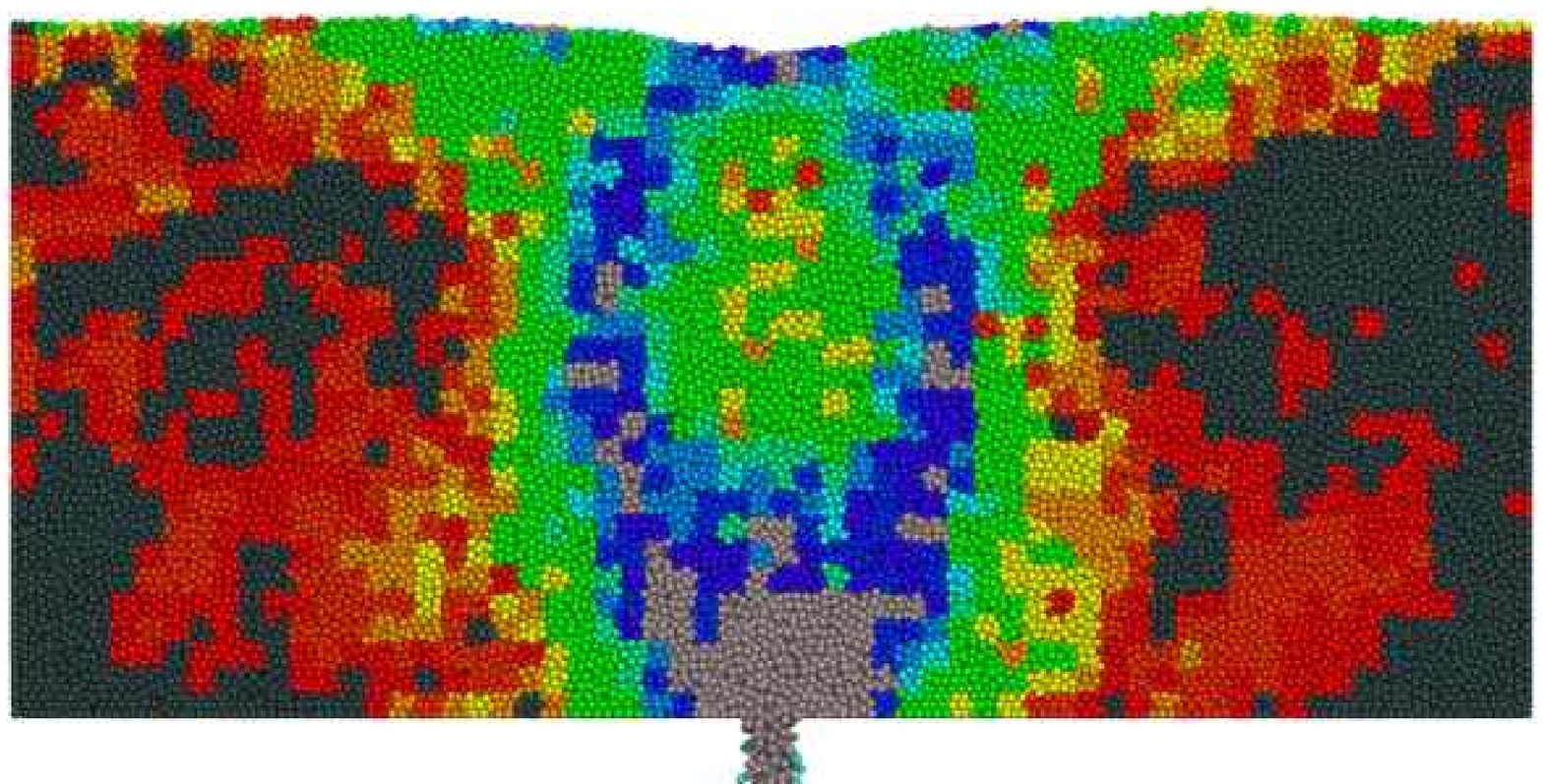, width=5.1in, clip}
\caption{The plastic shearing rate $d^p$, expressed in units of
  $\sqrt{g/d}$, plotted in logarithm form to accentuate small
  features. (a) The elasto-plastic solution: Note the intricate
  pattern of shear bands that fill the region between the two long
  shearing arms.  The long-time behavior has the bands fall down
  one-by-one from the larger shearing arms. (b) The DEM solution c/o
  Rycroft \cite{rycroft09}: Similar to the elasto-plastic except
  blurred out by the box-averaging and the effects of non-local
  ``diffusion'' that are ignored by first-order
  elasto-plasticity. Both plots use the same color
  scale.}\label{log_dp}
\end{figure}

\begin{figure}
\begin{center}
(a) \ \ \ \ \epsfig{file=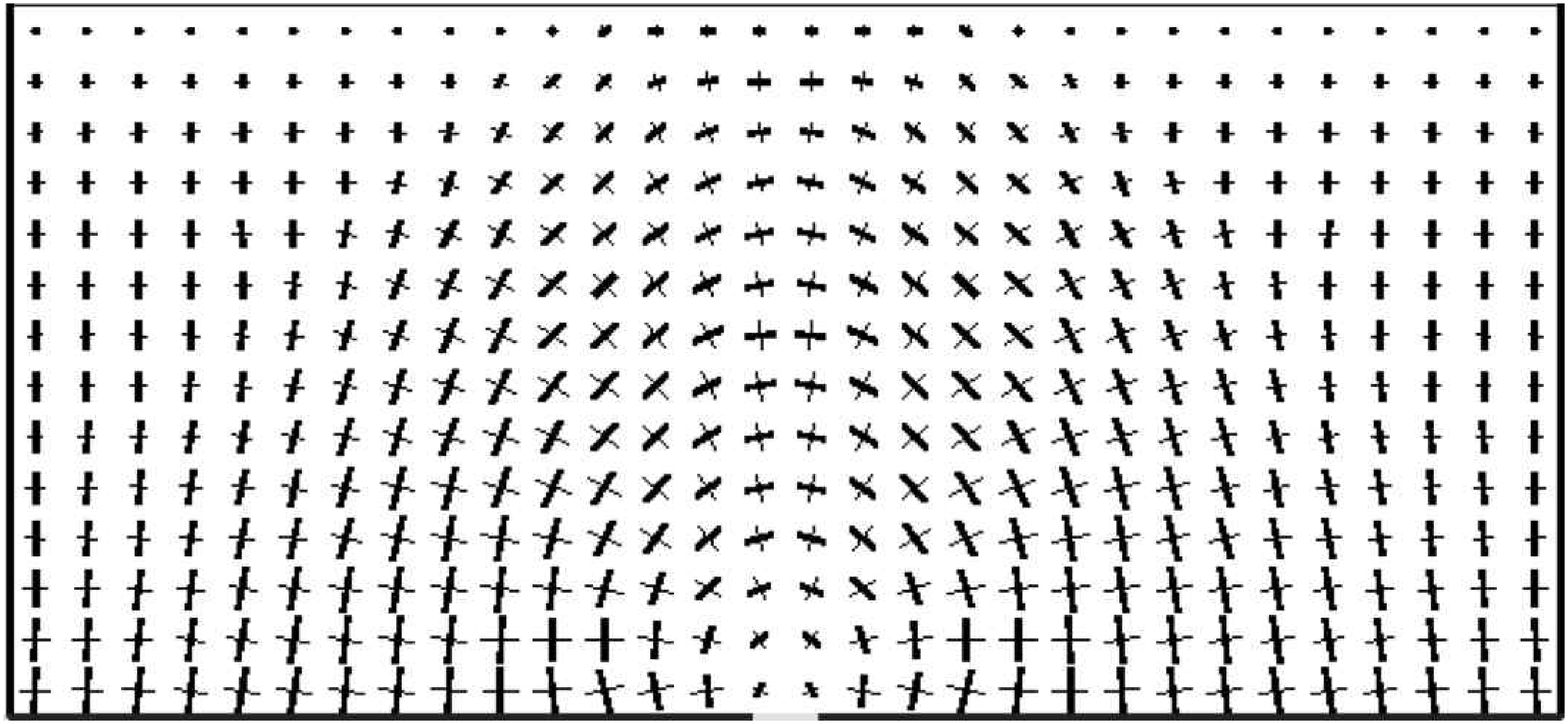, width=5.1in ,clip}

\bigskip
\bigskip
\bigskip

(b) \ \ \ \ \epsfig{file=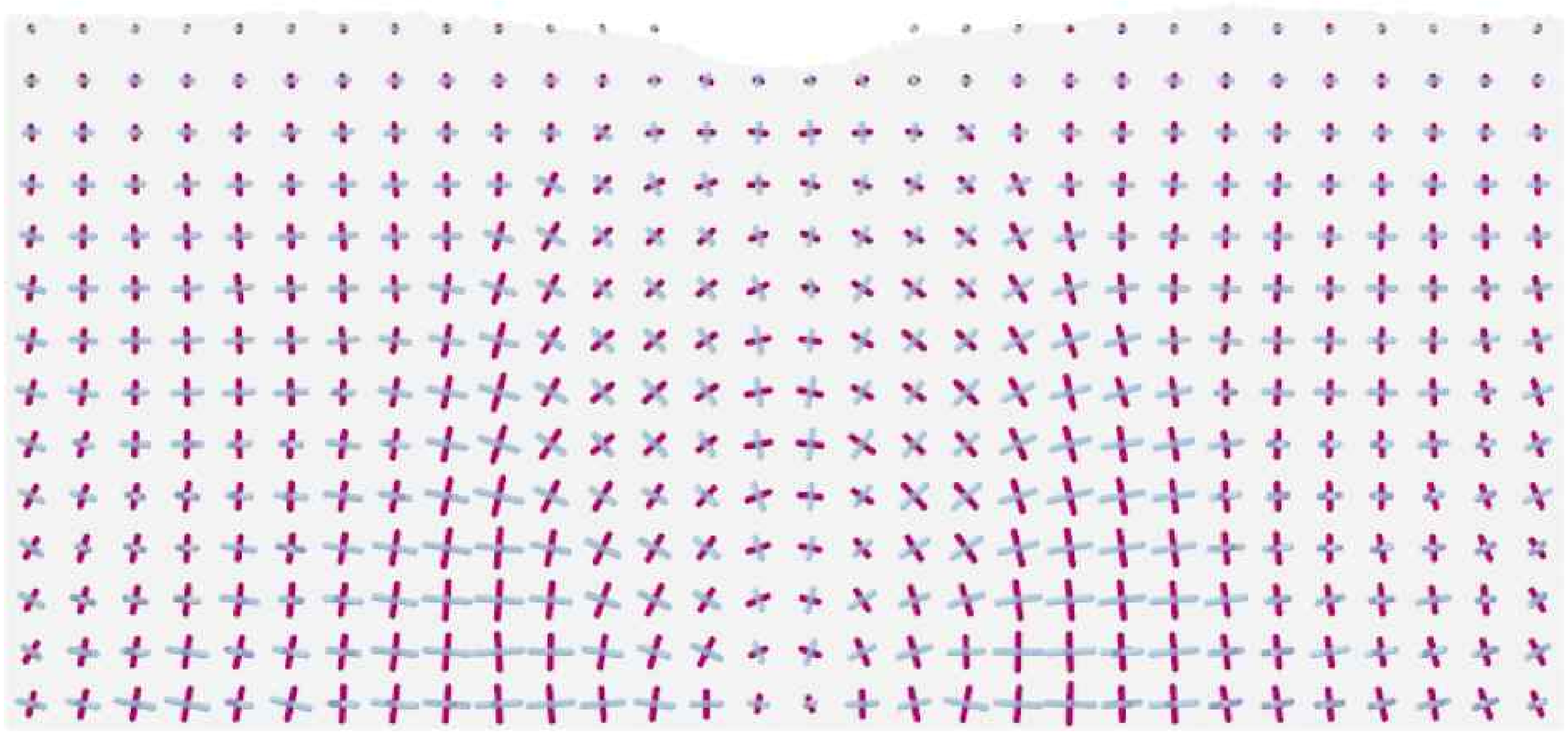, width=5.1in, clip}
\caption{The instantaneous deviatoric principal stress directions
  plotted as crosshairs with lengths corresponding to the associated
  deviatoric stress eigenvalues.  (a) The elasto-plastic solution: The
  thicker of the lines corresponds to the major principal stress
  direction. (b) The DEM solution c/o Rycroft \cite{rycroft09}:
  Major principal stress in purple, minor (and intermediate where
  visible) in blue.}\label{princ_silo}
\end{center}
\end{figure}

\begin{figure}
(a)\epsfig{file=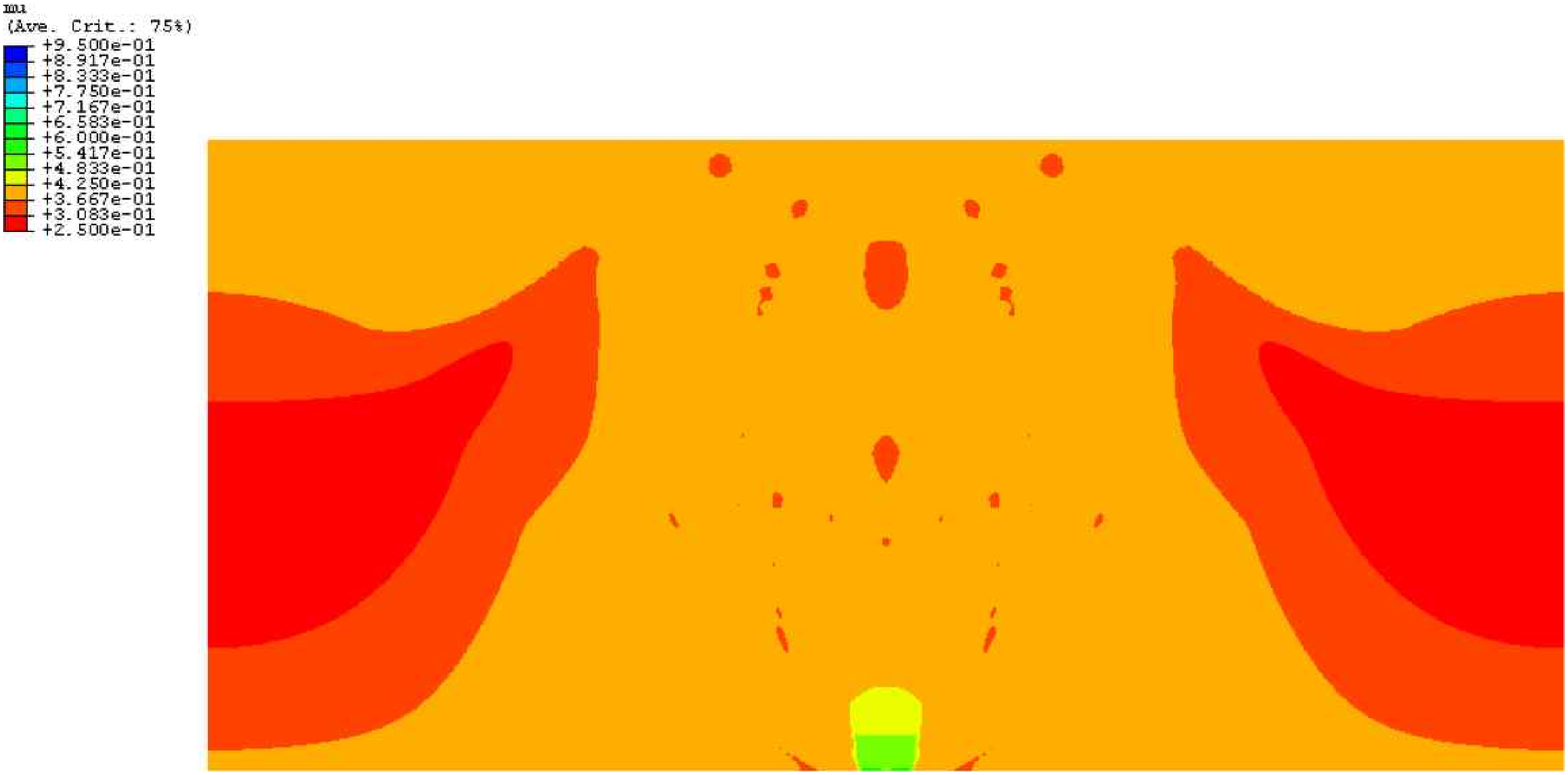, width=5.7in ,clip}

(b)\hspace{.64in}\epsfig{file=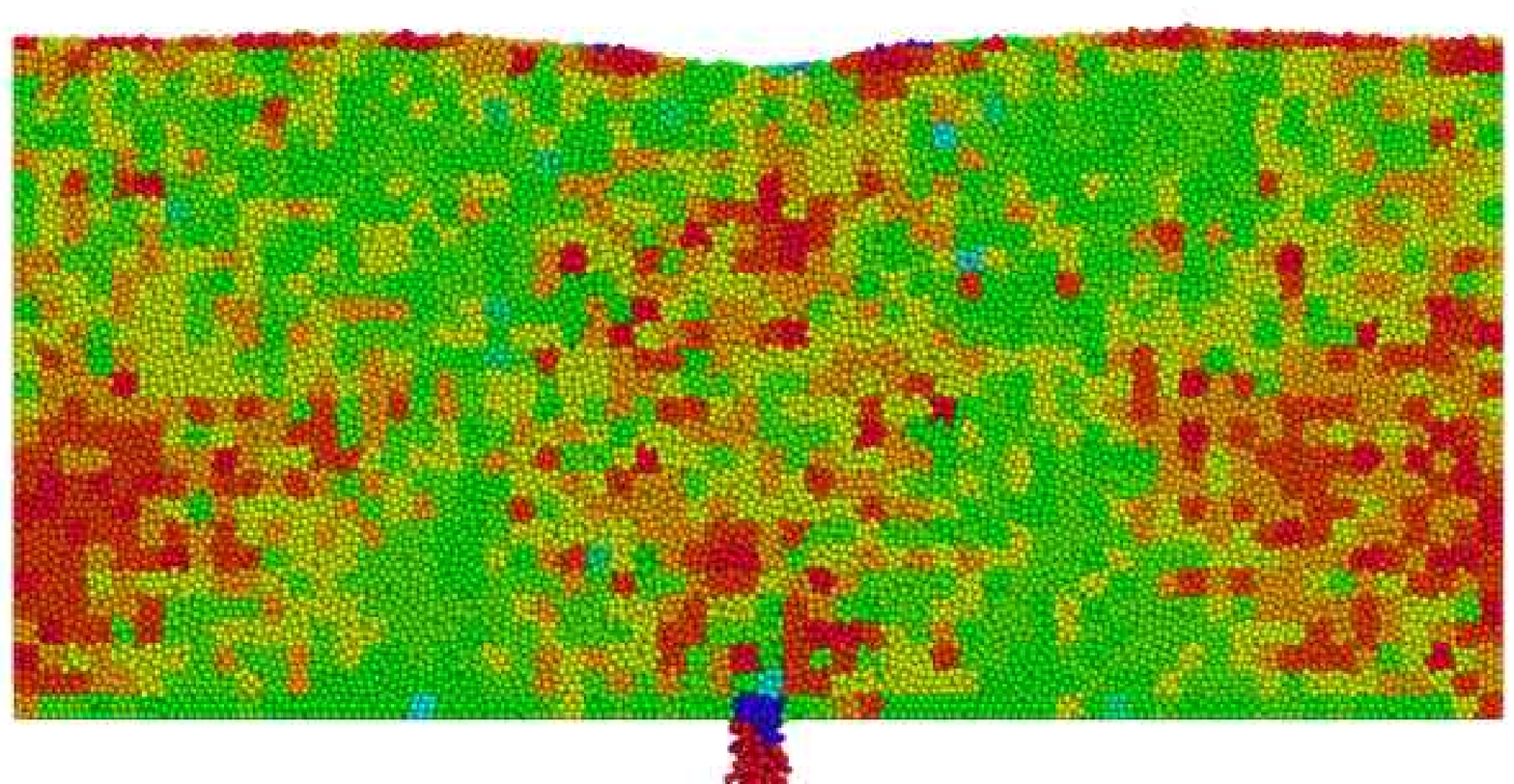, width=5.1in, clip}
\caption{The instantaneous stress ratio $\mu$ during fully developed
  flow. (a) The elasto-plastic solution. (b) The DEM solution c/o Rycroft \cite{rycroft09}. Both
  plots use the same color scale.}\label{mu_silo}
\end{figure}

The problem solved was as follows: From $t=0-5\times 10^{-5}$s,
gravity is gradually turned on while constraining the nodes along the
opening from moving in the $y$ direction in order to model a closed
orifice.  Then, from $t=0.55-5.00\times 10^{-4}$s, the orifice is
gradually opened by ramping up the outflow rate from zero to
$Q_{\text{out}}/2$.  The simulation is then left to flow until a total
time of $t=10^{-3}$s has been reached.

This environment does not possess a steady-state since it is not an
Eulerian boundary-value problem at the free surface.  Instead,
patternistic behavior eventually occurs, which signifies that
transients have finished passing--- starting at approximately
$t=5\times 10^{-4}$s, the velocity and $d^p$ fields appear to
fluctuate regularly. 

The first direct comparison that should be made is between the
elasto-plastic and DEM flow profiles. To represent fully-developed
mean behavior, Rycroft's flow data was averaged over 100 frames,
during a period of what appears to be transient-free flow.  Similarly,
the elasto-plastic flow was also averaged over many frames of
fully-developed motion.  To improve the validity of this average, the
model was run an extra $5\times 10^{-4}$s longer and the time average
was performed over the range $t=0.5-1.5 \times 10^{-4}$s, comprising
127 frames. Figure \ref{silo_vel} displays the comparison.  Overall
the agreement is sufficient.  The particular way in which the peak in
the downward velocity component broadens as height increases is well
captured by the model.  Once again, as is now a common theme, the
elasto-plastic downward velocity appears to change more rapidly in
space.  Non-local effects such as diffusion could smooth out these
sharper variations and possibly improve the agreement. Observing
figure \ref{silo_vel}(c), the DEM and elasto-plastic trajectories
agree well, especially below $y\approx 40d$. The differences in the
upper silo trajectories could stem from the fact that in the DEM
simulation, the top free surface lowered notably by the time steady
flow could be imaged, whereas the elasto-plastic simulation reached
this point before any noticeable drop of the top free surface (due to
the artificial density reduction).

To compare instantaneous behavior, we observe snapshots of the
shearing rate profile. As shown in figure \ref{log_dp}, well-developed
elasto-plastic behavior involves two long ``arms'' of shearing that
extend from the edges of the orifice to the top surface.  From those
arms, narrower, weaker shear bands (note the log-scale) continually
drop down, creating an intersecting ``mesh''. The DEM solution does
look similar, with two dominant arms of shearing. Note as well that
near the top of the silo, both plots show the shearing spreads out
approaching the free surface.  The most obvious difference, as has
been previously noted, is that the elasto-plastic solution has sharp
flow cutoffs whereas the shear rate always gradually tails off in the
DEM.  To reiterate, the model lacks non-local quasi-static terms so
flow cutoffs are to be expected. Even if narrow bands did exist in the
DEM, the box-average being performed would most assuredly obscure
them.

Moving on to the stresses, we first check the principal stresses and
directions during fully-developed flow. As is evident in figure
\ref{princ_silo}, the principal stress orientations predicted by the
elasto-plastic model closely match those of the DEM.  Both show the
major principal stresses forming ``arches'' about the orifice and,
moving away from the opening, the principal stress chains transition
to becoming more vertical.  In the DEM, it is apparent that the major
principal stresses adjacent to the side walls have a slightly tilted
orientation, whereas those of the elasto-plastic solution remain
almost perfectly vertical.  This is entirely due to the chosen
boundary conditions.  The DEM utilized frictional sidewalls and the
tilt indicates the walls are exerting an upward shear on the material.
This wall shear can be traced back to the filling process, where
pouring causes an ``active wall state'' that resists the downward
motion of the grains.  On the contrary, the side walls in the
elasto-plastic simulation are frictionless.  A side wall friction may
be included in the future to better model this minor effect.

For a more direct comparison of the relative sizes of the stress
invariants, see figure \ref{mu_silo}.  Qualitatively, the two
solutions show similar spatial changes.  Both show a region of lower
$\mu$ that swoops up from the lower side walls together with scattered
minima of $\mu$ in the upper-middle region of the silo.
Quantitatively, it appears the $\mu$ values in the flowing zone are
lower for the elasto-plastic solution. A probable explanation for this
could be that the grains in the DEM simulation have a different
surface roughness than the glass beads of \cite{jop05} and should in
actuality have higher $\mu_s$ and $\mu_2$ values. Recall that our
material parameters were not extracted from the DEM data, but come
from two \emph{different} papers. In \cite{dacruz05} it was
demonstrated that increasing the surface roughness of grains causes
$\mu_s$ to increase.  If the elasto-plastic model were implemented
with a larger friction constants, the $\mu$ profile would indeed
increase in the flowing regions, as larger $\mu$ would be needed to
invoke the same plastic shear rate.

\section{Conclusion and future directions}

This work has demonstrated a highly general, 3D granular continuum
model that unifies the recent results in granular elasticity and
plasticity. The unification follows a rigorous $\vec{F}^e\vec{F}^p$
decomposition, with elastic and plastic response combined using
finite-deformation elasto-plasticity. The model can be used to predict
flows uniquely in any environment with mechanically well-posed
boundary conditions and/or body forces.  The model was implemented as
a user-material in ABAQUS/Explicit and tested in three unrelated
geometries. With no fitting, it appears to give qualitative, and in
some case quantitative, predictions for both the stress and velocity
field in arbitrary granular flow geometries. Even so, there are a few
clear avenues of future work with regard to improving the current
model.

\subsection{Quasi-static non-locality}

The most glaring effect absent from the model is that it cannot
account for ``blurring'' in the flow fields.  As described in depth in
sections \ref{regimes1} and \ref{regimes2}, when the normalized shear
rate decreases low enough, the rheology is no longer determinable from
a simple relation of the form $I=g^{-1}(\mu)$.  

One would speculate a more complete form has higher order gradients in
stress, flow rate, and/or state parameters balanced by some additional
grain-level length-scale. In regions of moderate flow rate, these
spatially second order effects should be dominated by the local
rheology. This is clear from the above tests, which show the model
does indeed perform better where the flow is faster. But equally clear
is the fact that these terms cannot be ignored near static yield.
Section \ref{regimes2} lists several possible theories to describe
non-local behavior, but how each candidate would fit theoretically
within the current model remains to be studied.

By neglecting quasi-static non-locality, the model is tacitly
assuming that the material is capable of forming clean
solid-like/fluid-like interfaces.  This is almost always an
idealization that is unrealized, as particles in a random packing
rarely have the geometric ability to assemble in a fashion that avoids
overlap with the predicted interface. A particle on the interface,
being unable to both shear with the fluid and remain static with the
solid, instead transmits some of the shearing behavior from the
fluid-like zone into the solid zone, thereby explaining the gradual
tails we see in actual granular flow profiles.  Under particular
circumstances, however, granular flow can be made to segregate clearly
into flowing and completely static zones.  Thompson and Grest
\cite{thompson91} have shown that a monodisperse 2D disk assembly
undergoing horizontal planar shear with downward gravity does indeed
have a flow cutoff with zero flow occurring beneath a shear band at the
top.  This behavior is a rarity brought on by the fact that solid-like
material in this geometry can and does form a hexagonal crystal and
the horizontal fluid/solid interface happens to align perfectly with a
crystal plane.

\subsection{Dilation}
The present model avoids all inelastic dilation.  Though an argument for
why this is acceptable for our current purposes is presented in
section \ref{jop}, certain benefits would come with properly
accounting for the small amount of dilation that occurs in dense flow.

Inelastic dilation comes in two primary forms: plastic and gas-like.
Shear and shearing dilation, as described in section \ref{regimes1},
are prototypical plastic mechanisms.  Gas-like dilation can occur in
particularly energetic surroundings, even below the plastic yield
criterion--- for example, consider shaking a box of rubber spheres
under uniform pressure. Granted, gas-like behavior is outside the
dense flow regime we currently study.  Even so, it is possible for a
small region of an otherwise dense flow to become gas-like, and
consequently a means of dealing with this behavior is desired. Recall
this was a crucial issue in the silo geometry, which is almost
entirely dense except within the small free-fall arch that encompasses
the opening. Gas-like effects may require a granular temperature and
heat flux to express the internal pressure.

Several models for plastic dilation have been proposed (see section
\ref{jop}), but a direct 3D discrete element study would be ideal for
quantifying the precise dilatational dependences. One might
hypothesize, based on equation \ref{I_Phi}, that the plastic dilation
has a form
\[\frac{d \eta}{d \gamma^p}=\frac{\dot{\eta}}{\tilde{d}^p(\vec{M})}=A(\eta) B(\Phi(I)-0.63 e^{-\eta})\]
where $\eta=\log(\text{det} \vec{F}^p)$ measures plastic dilation,
$\gamma^p$ is a plastic shear strain,
$I=\tilde{d}^p(\vec{M})/\left(\frac{1}{d}\sqrt{\frac{P}{\rho_s}}\right)$ is the
inertial number, and the functions $A$ and $B$ are empirical, with
$B=1$ when the magnitude of its argument is large and $B\rightarrow0$
as the input goes to zero.  The above states that a material
originally at random close packing ($\Phi=0.63$) obeys Bagnold type
dependences at steady flow. While the flow is unsteady, it dilates
according to $A$ and then ceases dilation as the Bagnold relationship
is approached. A relation of this form appears to agree with results
of Rycroft $\etal$ \cite{rycroft09}, and collaborative efforts with Rycroft
are underway to quantify this evolution law.  

Once a form for the plastic dilation has been verified, it would serve
to enhance the computation of elasto-statics, where the moduli are
known to vary with the packing fraction, as well as give meaningful
packing fraction data throughout.  However, the technique of
artificial density reduction may no longer be valid, since the steady
packing fraction, while only a few percent different than at the
start, may take more time to develop than is typically allotted in a
simulation run.

\subsection{Flow condition}
The plastic flow rule of Jop \etal asserts the codirectionality
condition. A direct test of the flow condition should be performed to
verify whether this is indeed the best candidate. Such a test may be
forthcoming with recent DEM data on large conical hopper flow.  Some
models have had success utilizing double-shearing conditions
\cite{anand00}, and Rycroft's silo flow data has shown some deviation
from codirectional flow.  Even so, codirectionality has fit the
current needs and appears to be sufficient to predict basic flow
behavior. A fine-tuned flow condition would be essential to
model highly asymmetric 3D flows.

\section*{Acknowledgements}

The author gratefully acknowledges Lallit Anand and the MIT Solid
Mechanics Group for helpful discussion and access to the ABAQUS
software package. The author also thanks Chris H. Rycroft for access
to discrete simulation data and acknowledges Martin Z. Bazant for
advice. This work was supported by the NDSEG and NSF GRFP
fellowship programs.


\begin{thebibliography}{10}

\bibitem{anand00}
L.~Anand and C.~Gu.
\newblock Granular materials: constitutive equations and strain localization.
\newblock {\em J. Mech. Phys. Solids}, 28:1701, 2000.

\bibitem{anand04}
L.~Anand and C.~Su.
\newblock A theory for amorphous viscoplastic materials undergoing finite
  deformations, with application to metallic glasses.
\newblock {\em J. Mech. Phys. Solids}, 53:1362--1396, 2005.

\bibitem{aranson01}
I.~S. Aranson and L.~S. Tsimring.
\newblock Continuum description of avalanches in granular media.
\newblock {\em Phys. Rev. E}, 64:020301, 2001.

\bibitem{aranson02}
I.~S. Aranson and L.~S. Tsimring.
\newblock Continuum theory of partially fluidized granular flows.
\newblock {\em Phys. Rev. E}, 65:061303, 2002.

\bibitem{aranson06}
I.~S. Aranson and L.~S. Tsimring.
\newblock Patterns and collective behavior in granular media: Theoretical
  concepts.
\newblock {\em Rev. Mod. Phys.}, 78:641--692, 2006.

\bibitem{bagnold54}
R.~A. Bagnold.
\newblock Experiments on a gravity free dispersion of large solid spheres in a
  newtonian fluid under shear.
\newblock {\em Proc. Roy. Soc. London Ser. A}, 225, 1954.

\bibitem{alltheories}
N.J. Balmforth and A.~Provenzale.
\newblock Patterns of dirt.
\newblock {\em Geomorph. Fluid Mech.}, 582:164--187, 2001.

\bibitem{bazant06}
M.~Z. Bazant.
\newblock The spot model for random-packing dynamics.
\newblock {\em Mechanics of Materials}, 38:717--731, 2006.

\bibitem{behringer05}
R.~P. Behringer.
\newblock Contact force measurements and stress-induced anisotropy in granular
  materials.
\newblock {\em Nature}, 435:1079--1082, 2005.

\bibitem{bocquet02}
L.~Bocquet, W.~Losert, D.~Schalk, T.~C. Lubensky, and J.~P. Gollub.
\newblock Granular shear flow dynamics and forces: Experiment and continuum
  theory.
\newblock {\em Phys. Rev. E}, 65:011307, 2002.

\bibitem{bouchaud95}
J.-P. Bouchaud, M.~E. Cates, and P.~Claudin.
\newblock Stress distribution in granular media and nonlinear wave equation.
\newblock {\em J. Phys.}, 5:639--656, 1995.

\bibitem{bouchaud94}
J.-P. Bouchaud, M.~E. Cates, J.~R. Prakash, and S.~F. Edwards.
\newblock A model for the dynamics of sandpile surfaces.
\newblock {\em J. Phys. I (France)}, 4:1383, 1994.

\bibitem{bouchaud95a}
J.-P. Bouchaud, M.~E. Cates, J.~R. Prakash, and S.~F. Edwards.
\newblock Hysteresis and metastability in a continuum sandpile model.
\newblock {\em Phys. Rev. Lett.}, 74:1982, 1995.

\bibitem{boutreux99}
T.~Boutreux, H.~A. Makse, and P.-G. de~Gennes.
\newblock Surface flows of granular mixtures.
\newblock {\em Euro. Phys. J. B}, 9:105--115, 1999.

\bibitem{boutreux98}
T.~Boutreux, E.~Rapha\"el, and P.-G. de~Gennes.
\newblock Surface flows of granular materials: A modified picture for thick
  avalanches.
\newblock {\em Phys. Rev. E}, 58:4692--4700, 1998.

\bibitem{choi05}
J.~Choi, A.~Kudrolli, and M.~Z. Bazant.
\newblock Velocity profile of gravity-driven dense granular flow.
\newblock {\em J. Phys.: Condensed Matter}, 17:S2533--S2548, 2005.

\bibitem{dacruz05}
F.~da~Cruz, S.~Emam, M.~Prochnow, J.~Roux, and F.~Chevoir.
\newblock Rheophysics of dense granular materials: Discrete simulation of plane
  shear flows.
\newblock {\em Phys. Rev. E.}, 72:021309, 2005.

\bibitem{degennes99}
P.~G. de~Gennes.
\newblock Granular matter: a tentative view.
\newblock {\em Rev. Mod. Phys.}, 71:S374--S382, 1999.

\bibitem{depken07}
M.~Depken, J.~B. Lechman, M.~van Hecke, W.~van Saarloos, and G.~S. Grest.
\newblock Stresses in smooth flows of dense granular media.
\newblock {\em Europhys. Lett.}, 78:58001, 2007.

\bibitem{digby81}
P.~J. Digby.
\newblock The effective elastic modulus of porous granular rocks.
\newblock {\em J. Appl. Mech.}, 48:803, 1981.

\bibitem{Dre}
A.~Drescher.
\newblock {\em Analytical Methods in Bin-Load Analysis}.
\newblock Elsevier, 1991.

\bibitem{duffy57}
J.~Duffy and R.~D. Mindlin.
\newblock Stress--strain relation and vibrations of granular medium.
\newblock {\em J. Appl. Mech.}, 24:585, 1957.

\bibitem{edwards91}
S.~F. Edwards.
\newblock In Blackman and Taguena, editors, {\em Disorder in Condensed Matter
  Physics}. Oxford, 1991.

\bibitem{edwards01}
S.~F. Edwards and D.~V. Grinev.
\newblock Granular media as a physics problem.
\newblock {\em Advances in Complex Systems}, 4:451--467, 2001.
\newblock (also reprinted in Ref.~\protect\cite{halsey02}).

\bibitem{ertas02}
D.~Erta{\c s} and T.~C. Halsey.
\newblock Granular gravitational collapse and chute flow.
\newblock {\em Europhys. Lett.}, 60:931--937, 2002.

\bibitem{evesque98}
P.~Evesque and P.~G. de~Gennes.
\newblock On the statics of silos.
\newblock {\em C. R. Acad. Sci. (PARIS), Ser. II}, 326:761, 1998.

\bibitem{falk98}
M.~L. Falk and J.~S. Langer.
\newblock Dynamics of viscoplastic deformation in amorphous solids.
\newblock {\em Phys. Rev. E}, 57:7192--7205, 1998.

\bibitem{geng01}
J.~Geng, D.~Howell, E.~Longhi, R.~P. Behringer, G.~Reydellet, L.~Vanel,
  E.~Cl\`{e}ment, and S.~Luding.
\newblock Footprints in sand: The response of a granular material to local
  perturbations.
\newblock {\em Phys. Rev. Lett.}, 87:035506, 2001.

\bibitem{glasser01}
B.~J. Glasser and I.~Goldhirsch.
\newblock Scale dependence, correlations, and fluctuations of stresses in rapid
  granular flows.
\newblock {\em Phys. Fluids}, 13:407, 2001.

\bibitem{goddard90}
J.~D. Goddard.
\newblock Nonlinear elasticity and pressure-dependent wave speeds in granular
  media.
\newblock {\em Proc. R. Soc. London, Ser. A}, 430:105, 1990.

\bibitem{goldenberg06}
C.~Goldenberg, A.~P.~F. Atman, P.~Claudin, G.~Combe, and I.~Goldhirsch.
\newblock Scale separation in granular packings: Stress plateaus and
  fluctuations.
\newblock {\em Phys. Rev. Lett.}, 96:168001, 2006.

\bibitem{goldenberg05}
C.~Goldenberg and I.~Goldhirsch.
\newblock Friction enhances elasticity in granular solids.
\newblock {\em Nature}, 435:188--191, 2005.

\bibitem{gurtin05a}
M.~E. Gurtin and L.~Anand.
\newblock A theory of strain-gradient plasticity for isotropic, plastically
  irrotational materials. part i: Small deformations.
\newblock {\em J. Mech. Phys. Solids}, 53:1642--1649, 2005.

\bibitem{gurtin05b}
M.~E. Gurtin and L.~Anand.
\newblock A theory of strain-gradient plasticity for isotropic, plastically
  irrotational materials. part ii: Finite deformations.
\newblock {\em J. Mech. Phys. Solids}, 21:2297--2318, 2005.

\bibitem{gurtin81}
Morton~E. Gurtin.
\newblock {\em An Introduction to Continuum Mechanics}.
\newblock Academic Press, 1981.

\bibitem{halsey02}
T.~Halsey and A.~Mehta, editors.
\newblock {\em Challenges in Granular Physics}.
\newblock World Scientific, 2002.

\bibitem{hashiguchi07}
K.~Hashiguchi and S.~Tsutsumi.
\newblock Gradient plasticity with the tangential-subloading surface model and
  the prediction of shear-band thickness of granular materials.
\newblock {\em Int. J. Plasticity}, 23:767--797, 2007.

\bibitem{hattamleh04}
O.~Al Hattamleh, B.~Muhunthan, and H.~M. Zbib.
\newblock Gradient plasticity modelling of strain localization in granular
  materials.
\newblock {\em Int. J. Numer. Anal. Meth. Geomech.}, 28:465--481, 2004.

\bibitem{hill}
R.~Hill.
\newblock {\em The Mathematical Theory of Plasticity}.
\newblock Oxford at the Clarendon Press, 1950.

\bibitem{iordanoff04}
I.~Iordanoff and M.~M. Khonsari.
\newblock Granular lubrication: toward an understanding between kinetic and
  fluid regime.
\newblock {\em ASME J. Tribol.}, 126:137--145.

\bibitem{jaeger96}
H.~M. Jaeger, S.~R. Nagel, and R.~P. Behringer.
\newblock Granular solids, liquids, and gases.
\newblock {\em Rev. Mod. Phys.}, 68:1259--1273, 1996.

\bibitem{jiang03}
Y.~Jiang and M.~Liu.
\newblock Granular elasticity without the coulomb condition.
\newblock {\em Phys. Rev. Lett.}, 91:144301, 2003.

\bibitem{jop05}
P.~Jop, Y.~Forterre, and O.~Pouliquen.
\newblock Crucial role of side walls for granular surface flows: consequences
  for the rheology.
\newblock {\em J. Fluid, Mech.}, 541:21--50, 2005.

\bibitem{jop06}
P.~Jop, Y.~Forterre, and O.~Pouliquen.
\newblock A constitutive law for dense granular flows.
\newblock {\em Nature}, 441:727, 2006.

\bibitem{kadanoff99}
L.~P. Kadanoff.
\newblock Built upon sand: Theoretical ideas inspired by the flow of granular
  materials.
\newblock {\em Rev. Mod. Phys.}, 71:435--444, 1999.

\bibitem{kamrin_thesis}
K.~Kamrin.
\newblock {\em Stochastic and Deterministic Models for Dense Granular Flow}.
\newblock PhD thesis, Massachusetts Institute of Technology, 2008.

\bibitem{kamrin07a}
K.~Kamrin and M.~Z. Bazant.
\newblock Stochastic flow rule for granular materials.
\newblock {\em Phys. Rev. E}, 75:041301, 2007.

\bibitem{kamrin07b}
K.~Kamrin, C.~H. Rycroft, and M.~Z. Bazant.
\newblock The stochastic flow rule: A multi--scale model for granular
  plasticity.
\newblock {\em Modelling Simul. Mater. Sci. Eng.}, 15:S449, 2007.

\bibitem{koval_thesis}
Georg Koval.
\newblock {\em Compotement d'interface des mat\'{e}riaux granulaires}.
\newblock PhD thesis, \'{E}cole Nationale des Ponts et Chauss\'{e}es, 2008.

\bibitem{kroner60}
E.~Kr\"{o}ner.
\newblock Allgemeine kontinuumstheorie der versetzungen und eigenspannungen.
\newblock {\em Arch. Ration. Mech. Anal.}, 4:273--334, 1960.

\bibitem{kuwano02}
R.~Kuwano and R.~J. Jardine.
\newblock On the applicability of cross-anisotropic elasticity to granular
  materials at very small strains.
\newblock {\em G\'{e}otechnique}, 52:10:727--749, 2002.

\bibitem{lee69}
E.~H. Lee.
\newblock Elastic plastic deformation at finite strain.
\newblock {\em J. Appl. Mech.}, 36:1--6, 1969.

\bibitem{lemaitre02c}
Ana\"el Lema\^itre.
\newblock Origin of a repose angle: Kinetics of rearrangements for granular
  materials.
\newblock {\em Phys. Rev. Lett.}, 89:064303, 2002.

\bibitem{lemaitre02}
Ana\"el Lema\^itre.
\newblock Rearrangements and dilatency for sheared dense materials.
\newblock {\em Phys. Rev. Lett.}, 89:195503, 2002.

\bibitem{lit58}
J.~Litwiniszyn.
\newblock Statistical methods in the mechanics of granular bodies.
\newblock {\em Rheol. Acta}, 2/3:146, 1958.

\bibitem{lois05}
G.~Lois, A.~Lemaitre, and J.~M. Carlson.
\newblock Numerical tests of constitutive laws for dense granular flows.
\newblock {\em Phys. Rev. E.}, 72:051303, 2005.

\bibitem{losert00}
W.~Losert, L.~Bocquet, T.~C. Lubensky, and J.~P. Gollub.
\newblock Particle dynamics in sheared granular matter.
\newblock {\em Phys. Rev. Lett.}, 85:1428--1431, 2000.

\bibitem{makse04}
H.~A. Makse, N.~Gland, D.~L. Johnson, and L.~Schwartz.
\newblock Granular packings: Nonlinear elasticity, sound propagation, and
  collective relaxation dynamics.
\newblock {\em Phys. Rev. E}, 70:061302, 2004.

\bibitem{midi04}
G.~D.~R. Midi.
\newblock On dense granular flows.
\newblock {\em Euro. Phys. Journ. E.}, 14:341--365, 2004.

\bibitem{mills99}
P.~Mills, D.~Loggia, and M.~Tixier.
\newblock Model for a stationary dense granular flow along an inclined wall.
\newblock {\em Europhys. Lett.}, 45:733--738, 1999.

\bibitem{mindlin53}
R.~D. Mindlin and H.~Deresiewicz.
\newblock Elastic spheres in contact under varying oblique forces.
\newblock {\em J. Appl. Mech.}, 20:327--244, 1953.

\bibitem{mullins72}
J.~Mullins.
\newblock Stochastic theory of particle flow under gravity.
\newblock {\em J. Appl. Phys.}, 43:665, 1972.

\bibitem{nedderman}
R.~M. Nedderman.
\newblock {\em Statics and Kinematics of Granular Materials}.
\newblock Cambridge University Press, 1992.

\bibitem{nedderman79}
R.~M. Nedderman and U.~T\"uz\"un.
\newblock Kinematic model for the flow of granular materials.
\newblock {\em Powder Technology}, 22:243, 1979.

\bibitem{norris97}
A.~N. Norris and D.~L. Johnson.
\newblock Nonlinear elasticity of granular media.
\newblock {\em J. Appl. Mech.}, 64:39, 1997.

\bibitem{ostoja06}
M.~Ostoja-Starzewski.
\newblock Material spatial randomness: From statistical to representative
  volume element.
\newblock {\em Prob. Eng. Mech.}, 21:112--132.

\bibitem{ostoja05}
M.~Ostoja-Starzewski.
\newblock Scale effects in plasticity of random media: status and challenges.
\newblock {\em Int. J. Plasticity}, 21:1119--1160, 2005.

\bibitem{pouliquen99}
O.~Pouliquen.
\newblock Scaling laws in granular flows down rough inclined planes.
\newblock {\em Phys. Fluids}, 11:542, 1999.

\bibitem{pouliquen01}
O.~Pouliquen, Y.~Forterre, and S.~Le Dizes.
\newblock Slow dense granular flows as a self-induced process.
\newblock {\em Advances in Complex Systems}, 4:441--450, 2001.
\newblock (also reprinted in Ref.~\protect\cite{halsey02}).

\bibitem{pouliquen96}
O.~Pouliquen and R.~Gutfraind.
\newblock Stress fluctuations and shear zones in quasistatic granular chute
  flows.
\newblock {\em Phys. Rev. E}, 53:552--561, 1996.

\bibitem{prochnow_thesis}
M.~Prochnow.
\newblock {\em Ecoulemnets denses de grains secs.}
\newblock PhD thesis, Ecole Nationale de Ponts et Chauss\'{e}es, Marne la
  Vall\'{e}e, France, 2002.

\bibitem{ragione08}
L.~La Ragione, V.~C. Prantil, and I.~Sharma.
\newblock A simplified model for inelastic behavior of an idealized granular
  material.
\newblock {\em Int. J. Plasticity}, 24:168--189, 2008.

\bibitem{rudnicki75}
J.~W. Rudnicki and J.~R. Rice.
\newblock Conditions for the localization of deformation in pressure-sensitive
  and dilatant materials.
\newblock {\em J. Mech. Phys. Solids}, 23:371, 1975.

\bibitem{rycroft_misc}
C.~H. Rycroft.
\newblock private communication.

\bibitem{rycroft06a}
C.~H. Rycroft, M.~Z. Bazant, G.~S. Grest, and J.~W. Landry.
\newblock Dynamics of random packings in granular flow.
\newblock {\em Physical Review E}, 73:051306, 2006.

\bibitem{rycroft09}
C.~H. Rycroft, K.~Kamrin, and M.~Z. Bazant.
\newblock Assessing continuum hypotheses in simulation of granular flow.
\newblock In press, DOI: 10.1016/j.jmps.2009.01.009, 2009.

\bibitem{samadani99}
A.~Samadani, A.~Pradhan, and A.~Kudrolli.
\newblock Size segregation of granular matter in silo discharges.
\newblock {\em Phys. Rev. E}, 60:7203--7209, 1999.

\bibitem{savage98}
S.~B. Savage.
\newblock Analyses of slow high-concentration flows of granular materials.
\newblock {\em J. Fluid Mech.}, 377:1, 1998.

\bibitem{wroth}
A.~Schoefield and P.~Wroth.
\newblock {\em Critical State Soil Mechanics}.
\newblock McGraw-Hill, 1968.

\bibitem{silbert01}
L.~E. Silbert, D.~Ertas, G.~S. Grest, T.~C. Halsey, D.~Levine, and S.~J.
  Plimpton.
\newblock Granular flow down an inclined plane: Bagnold scaling and rheology.
\newblock {\em Phys. Rev. E.}, 64:051302, 2001.

\bibitem{Sok}
V.~V. Sokolovskii.
\newblock {\em Statics of Granular Materials}.
\newblock Pergamon/Oxford, 1965.

\bibitem{spencer64}
A.~J.~M. Spencer.
\newblock A theory of the kinematics of ideal soils under plane strain
  conditions.
\newblock {\em J. Mech. Physics}, 12:337–351, 1964.

\bibitem{thompson91}
P.~A. Thompson and G.~S. Grest.
\newblock Granular flow - friction and the dilatancy transition.
\newblock {\em Phys Rev Lett}, 67:1751--1754, 1991.

\bibitem{thornton06}
C.~Thornton and L.~Zhang.
\newblock A numerical examination of shear banding and simple shear non-coaxial
  flow rules.
\newblock {\em Phil. Mag.}, 86:3425--3452, 2006.

\bibitem{tsai03}
J.-C. Tsai, G.~A. Voth, and J.~P. Gollub.
\newblock Internal granular dynamics, shear-induced crystallization, and
  compaction steps.
\newblock {\em Phys. Rev. Lett.}, 91:064301, 2003.

\bibitem{tsutsumi05}
S.~Tsutsumi and K.~Hashiguchi.
\newblock General non-proportional loading behavior of soils.
\newblock {\em Int. J. Plasticity}, 21:1941--1969, 2005.

\bibitem{tsutsumi08}
S.~Tsutsumi and K.~Kaneko.
\newblock Constitutive response of idealized granular media under the principal
  stress axes rotation.
\newblock {\em Int. J. Plasticity}, 24:1967--1989, 2008.

\bibitem{vardoulakis91}
I.~Vardoulakis and E.~C. Aifantis.
\newblock A gradient flow theory of plasticity for granular media.
\newblock {\em Acta Mechanica}, 87:197--217, 1991.

\bibitem{walton87}
K.~Walton.
\newblock The effective elastic moduli of a random packing of spheres.
\newblock {\em J. Mech. Phys. Solids}, 35:213, 1987.

\bibitem{zhu06}
H.~Zhu, M.~M. Mehrabadi, and M.~Massoudi.
\newblock Three-dimensional constitutive relations for granular materials based
  on the dilatant double shearing mechanism and the concept of fabric.
\newblock {\em Int. J. Plasticity}, 22:826--857, 2006.

\end{thebibliography}
\end{document}